\documentclass[5p, twocolumn]{elsarticle}

\pdfoutput=1
\usepackage{xspace}
\usepackage{blkarray}
\usepackage{multirow}
\usepackage{subfig}
\usepackage{graphicx}
\usepackage{epstopdf}
\usepackage{wrapfig}

\usepackage[table,xcdraw]{xcolor}

\usepackage{textcomp}
\usepackage{gensymb}
\usepackage{siunitx}
\usepackage{rotating}
\usepackage{multirow}
\usepackage{commath}
\usepackage{multicol}
\usepackage{multirow}
\usepackage{blkarray}
\usepackage{relsize}
\usepackage{booktabs}
\usepackage{array}
\usepackage{amsmath}
\usepackage{tabularx}
\usepackage[utf8]{inputenc}
\usepackage{listings}

\usepackage{soul}

\usepackage{amssymb}
\usepackage{pifont}
\usepackage[hyphens]{url}
\usepackage[hidelinks]{hyperref}
\hypersetup{breaklinks=true}

\usepackage{color}
\newcommand{\red}[1]{\textcolor{red}{#1}}

\begin{document}

\newcommand{\sysname}{\textsc{IoT-Keeper}\xspace}
\newcommand{\backend}{\textsc{Keeper Service}\xspace}
\newcommand{\frontend}{\textsc{Keeper Gateway}\xspace}
\newcommand{\controller}{\textsc{Keeper Controller}\xspace}
\newcommand{\rpi}{\textsc{R-PI}\xspace}

\newcommand{\classM}{malicious\xspace}
\newcommand{\classB}{benign\xspace}

\newcommand{\policies}{security policies\xspace}
\newcommand{\policy}{security policy\xspace}

\newcommand{\flow}{traffic flow\xspace}
\newcommand{\flows}{traffic flows\xspace}

\newcommand{\session}{traffic session\xspace}
\newcommand{\sessions}{traffic sessions\xspace}

\newcommand{\ofrule}{OF-rule\xspace}
\newcommand{\ovs}{OVS\xspace}

\newcommand{\wifi}{WiFi\xspace}

\newcommand{\psweep}{$N_{sweep}$\xspace}
\newcommand{\pscan}{$N_{scan}$\xspace}
\newcommand{\dt}{$D_{t}$\xspace}

\newcommand{\metadata}{network metadata\xspace}
\newcommand{\predictionvar}{device activity\xspace}

\begin{frontmatter}

    \title{IoT-KEEPER: Securing IoT Communications in Edge Networks}

    \author{Ibbad Hafeez\textsuperscript{$\dagger$}} \ead{ibbad.hafeez@helsinki.fi}
    \author{Markku Antikainen\textsuperscript{$\dagger,\ddagger$}} \ead{markku.antikainen@hiit.fi}
    \author{Aaron Yi Ding\textsuperscript{$\ast$}} \ead{aaron.ding@tudelft.nl}
    \author{Sasu Tarkoma\textsuperscript{$\dagger, \ddagger$}} \ead{sasu.tarkoma@helsinki.fi}

    \address{\textsuperscript{$\dagger$}University of Helsinki, Helsinki, Finland}
    \address{\textsuperscript{$\ddagger$}Helsinki Institute of Information Technology, Helsinki, Finland}
    \address{\textsuperscript{$\ast$}Delft University of Technology, Delft, Netherlands}

    \begin{abstract}
The increased popularity of IoT devices have made them lucrative targets for attackers. 
Due to insecure product development practices, these devices are often vulnerable even to very trivial attacks and can be easily compromised.
Due to the sheer number and heterogeneity of IoT devices, it is not possible to secure the IoT ecosystem using traditional endpoint and network security solutions. 
To address the challenges and requirements of securing IoT devices in edge networks, we present \sysname, which is a novel system capable of securing the network against any malicious activity, in real time. 
The proposed system uses a lightweight anomaly detection technique, to secure both device-to-device and device-to-infrastructure communications, while using limited resources available on the gateway. It uses unlabeled network data to distinguish between benign and malicious traffic patterns observed in the network.
A detailed evaluation, done with real world testbed, shows that \sysname detects any device generating malicious traffic with high accuracy ($\approx 0.982$) and low false positive rate ($\approx 0.01$). The results demonstrate that \sysname is lightweight, responsive and can effectively handle complex D2D interactions without requiring explicit attack signatures or sophisticated hardware.

    \end{abstract}


\end{frontmatter}

\section{Introduction}
\label{sec:intro}

IoT-enabled automation systems have opened homes and industrial environments to countless new threats~\cite{6475947, SICARI2015146}. 
There are several reasons for the sad state of IoT device security. IoT development teams often work without sufficient resources and under strict time constraints.
These factors make it tempting for development team to cut corners, for example, by re-using unverified code snippets, insecure third-party libraries and not following secure software development practices~\cite{Senrio:SingleFlaw, DLink:SingleFlaw, 6975580}. 
These, and several other factors, result in production of inherently vulnerable devices for consumer markets. 

The number of device specific exploits is constantly increasing due to growing number of IoT installations in small office, home office (SOHO), and enterprise networks.
The adversaries can also re-use existing exploits from PC-platforms against IoT devices running a stripped down version of Linux or Windows as device firmware. 
Moreover, a vast majority of the IoT devices are connected to SOHO networks with no security in place except for the network address translation (NAT), which is done on the gateway. On several occasions, attackers have been able to compromise these devices, installed deep inside SOHO networks, to launch extremely large scale attacks~\cite{TechRepublicIoTDDoS, ArborReport}. 

Due to prevalence of insecure IoT devices, network owners can no longer rely on the assumption that all devices in their network are well-behaving and trustworthy~\cite{IoT:InvisibleNetwork}. 
While this, to some extent, applies to every network, it is a particular concern in SOHO environments where the network owners do not have the know-how or resources to improve security. 
This, together with the fact that IoT devices are rarely updated~\cite{barcena2015insecurity}, makes it probable that some devices in the network will, eventually, get compromised by an attacker.

There are three key solutions commonly used to secure PC and mobile devices: \textit{software updates}, \textit{endpoint security solutions}, and \textit{network-based security solutions}. However, there are numerous reasons that these solutions cannot be used for securing IoT devices. 
Firstly, in most cases, there is limited, if any, product life-cycle support available for IoT devices, as the manufacturers do not provide regular firmware updates or security patches for these devices~\cite{7980167,Yu:2015:HTF:2834050.2834095}.
Secondly, it is not possible to develop effective endpoint security solutions, such as anti-malwares, due to lack of firmware support, software APIs and limited resources available on IoT devices. 

Securing network communications is a practical solution for securing IoT devices because these devices require constant network connectivity for their operations. It is also easier to gain network access to user devices, compared to physical access.
Traditional network security solutions offer limited support for securing IoT because these solutions mainly rely on traffic signatures for anomaly detection and it is practically infeasible to obtain enough labeled data from heterogeneous IoT devices, to generate these signatures. 
High deployment and operational costs of network security solutions is also a limiting factor in the use of existing network security solutions for securing SOHO and small enterprise networks. 

Various techniques have been proposed to detect anomalies in network traffic using machine learning~\cite{Gu:2008:BCA:1496711.1496721, DBLP:journals/corr/abs-1802-09089, DBLP:journals/corr/abs-1804-07474}. The applicability of these techniques depends on how accurately the classification model can capture given devices' benign network behavior and use to it identify malicious traffic produced by that device.
It is also challenging to keep these classification models up-to-date, as the devices' network behavior can change substantially due to firmware updates and configuration changes.

The aim of this work is to propose a system addressing the challenges and needs of securing IoT in SOHO and enterprise networks. Such a system should actively \textit{monitor} the network to identify and block any malicious traffic flows. 
It should maintain an up-to-date classification model detecting anomalies in network behavior of connected devices.
The system needs to be \textit{lightweight} and \textit{cost-efficient} to support wide-scale deployments using only limited hardware resources. It should be \textit{scalable} to support SOHO and enterprise scale deployments. Lastly, the system should ideally have high sensitivity, for identifying any malicious traffic, and low fall-out, to prevent false alarms.

In the light of these requirements, we propose \sysname, a novel system capable of classifying network traffic in real-time using semi-supervised machine learning techniques. \sysname monitors all traffic flows within and across the network. It identifies a devices' malicious behavior using the previous network activity of that device. In order to secure other devices in the network, \sysname uses \textit{adhoc overlay networks} to limit network access for any device generating malicious traffic. 

\sysname uses \textit{fuzzy C-Mean clustering} and \textit{fuzzy-interpolation scheme} to identify malicious network traffic. 
This technique is lightweight enough to allow deployments using single-board computers, for example Raspberry-PI, making \sysname cost-efficient and easy to deploy. 
Given the challenges of collecting labeled traffic data, the classification algorithm was developed to work with unlabeled traffic data.
This classification model can be represented as a set of rules, making it easier to share across multiple nodes, and improve scalability of the system. 

This work demonstrates how a simple yet efficient classification algorithm, when combined with sophisticated feature analysis, enables us to successfully classify network traffic, using only limited hardware resources. Consequently, \sysname can be realized using legacy hardware.

More specifically, our contributions are:
\begin{itemize}
\item Design and implementation of \sysname, a novel end-to-end solution, capable of blocking any \classM network activity
\item A simple and lightweight mechanism for dynamically enforcing network access control, at per-device, per-destination granularity, using \textit{adhoc overlay networks}.
\item Detailed study of individual features and their relative importance to formulate a set of most useful features for network traffic classification. 
\item A thorough investigation of \sysname performance using a real world testbed with 40+ devices. The evaluation results demonstrate that the proposed system is able to identify various types of network attacks, with high accuracy ($0.982$) and few false alarms ($0.01$), without any significant impact of user experience (latency increment $\approx 1.8\%$). 
\end{itemize}

\textit{Organization:} The rest of this paper is organized as follows. Section~\ref{sec:background} introduces the threat model and challenges faced in securing IoT ecosystems. Section~\ref{sec:sys_design} describes the design and architecture of proposed system in detail. In Section~\ref{sec:method}, we present the techniques developed for feature engineering and anomaly detection. Section~\ref{sec:dataset} describes the process of collecting dataset, used for training and evaluation of the proposed system. In Section~\ref{sec:evaluation}, we present the evaluation of \sysname, in terms of performance achieved for detecting anomalies, network throughput and system efficiency. In Section~\ref{sec:related_work}, we revise the current state of the art for securing IoT ecosystem. Section~\ref{sec:discussion} discusses possible shortcomings of the proposed solution. Finally, we present our conclusive statement about this work in Section~\ref{sec:conclusion}.

\section{Background}
\label{sec:background}

This paper refers to any network, where user devices are connected to get Internet access, as edge network, including SOHO and enterprise networks. A device with network connectivity and some sensing support can be referred to as \textit{connected} or \textit{IoT device}. This definition covers a wide range of devices, which can be divided into two categories:

\begin{itemize}
\item 
\textit{Single-purpose devices}: include resource constrained devices such as sensors, appliances, with limited or dedicated functionality. Users manage and interact with these devices using a smartphone or tablet device.

\item
\textit{Multi-purpose devices}: include high end devices such as smartphones and PCs, with better hardware resources. These devices support endpoint security solutions and device diagnostic tools.
Due to access to sensitive user information and much larger attack surface, numerous sophisticated attacks have been developed to compromise such devices.
\end{itemize}

For the sake of simplicity, the remainder of this paper will refer to any of these two types of devices as an \textit{IoT device} or simply \textit{device}. 
These devices require network connectivity for majority of their operations. We divide the network communications of these devices into two categories.

\begin{itemize}
\item
\textit{Device-to-Device (D2D) communications}: This category includes network communications among the devices connected to same network. These communications usually occur within a single broadcast domain.

\item
\textit{Device-to-Infrastructure (D2I) communications}: This category includes network communications between user devices and remote destinations. A remote destination can be any device or service operating in a different network or broadcast domain. 
\end{itemize}

Unless specified otherwise, the rest of paper refers to both these types of communication as \textit{network communication}.

Some single-purpose devices use low-power communication protocols, such as Bluetooth-LE (BLE) or Zigbee, to communicate with their respective \textit{IoT hub}. The hub then communicates with respective cloud service(s) via wired or wireless network. 
As a result, the network traffic to and from the hub gives a fairly accurate representation of the D2I communications of the IoT devices, connected to the hub.

We now discuss the threat model for IoT ecosystem and the challenges faced in securing the networks where these IoT devices are connected.

\subsection{Threat Model}
\label{sec:threat_model}

Edge networks typically contain a mixture of single-purpose and multi-purpose devices.
These networks are set up using a single gateway, to provide Internet access to all connected devices. The gateway offers basic security features such as, MAC/IP filtering. However, these features are not generally configured by users~\cite{Yu:2015:HTF:2834050.2834095}.
In case there is any intrusion detection system or firewall installed in the network, it only filters incoming and outgoing traffic, and treats all connected devices within the network as \emph{secure} and \emph{trusted}.

With IoT devices, this assumption about the trustworthiness of the devices does not hold, because it is fairly easy to exploit the vulnerable IoT devices, and thus gain access to the local network~\cite{CoffeeMachine:WiFi, SmartDoorbell:WiFi}. Owing to common network setups, once an adversary is within the network, it gets unwarranted access to perform any type of attack against other devices in the same network. These attacks can be categorized as:

\begin{enumerate}
\item 
\textbf{Network scanning}:
These attacks are used to recognize any TCP and UDP services that run at target hosts and to identify what kind of traffic filtering is done in the network. Network scanning can also be used to identify the firmware that is running on a target. These attacks are generally used to scan target nodes before launching dedicated attacks against the scanned targets. Commonly used variants of network scanning attacks include \textit{address-sweep}, \textit{port-sweep}, and \textit{port-scan} attacks. 

\item 
\textbf{Privilege escalation}:
Once the target is identified and scanned, the adversary tries to gain privileged access to it, in order to deploy malicious code. Many IoT devices use stripped down Linux as firmware and, therefore, attacker may try to invoke shell to gain root access. Attacker can also use factory-default credentials or device specific exploits to gain privileged access. Upon success, attacker is able to upload malicious code and perform desired state changes to compromise the target node.

\item 
\textbf{Man-In-The-Middle (MitM)}:
An adversary, connected to user network, can snoop-in on and interrupt all traffic in the network. It can use the communication patterns of legitimate user devices to conduct replay attacks. For example, an adversary can replay traffic intercepted from communication between smartphone and garage door sensor, to later open garage door without users knowledge. 
MitM attacks have serious security and privacy implications, as they can be used to steal user data and disrupt potentially critical devices~\cite{Gollakota:2011:THY:2018436.2018438}.

\item 
\textbf{Data theft}:
Health IoT, smart appliances, and similar devices collect a lot of data, which reveals a lot of information about their users. Typically, users do not have discrete control over how this data is collected and transmitted~\cite{Mortier:2016:PDM:3010079.3010082}. An attacker can compromise user devices to steal this data and use it for targeted attacks. 

\item 
\textbf{Botnets}:
Botnets are generally comprised of infected devices installed inside edge networks~\cite{6997492, Gu:2008:BCA:1496711.1496721}. 
These devices maintain normal state of their operations until a \textit{command \& control} server instructs them to launch an attack against specific target(s). Distributed Denial of Services (DDoS) attacks are a common example of how seemingly benign user devices are used to launch attacks at unprecedented scales~\cite{ArborReport, 203628}. 

\end{enumerate}

The goal of \sysname is to identify and block any variants of the aforementioned attacks in edge networks. More specifically, \sysname presumes that, for a given device, any deviation from its \classB behavior is motivated by malicious intent. Based on the type of attack the device was executing, network access restrictions are setup to limit the network activity of compromised device. This way, \sysname is able to block any compromised device from launching attacks against local or remote targets. 

\subsection{Challenges for Securing IoT}
\label{sec:challenges}
Conventional network and endpoint security solutions fall short in addressing the challenges of securing IoT ecosystem for a number of reasons. We now go through some of these.

Firstly, there is a huge diversity in IoT devices' firmwares, software stacks and APIs. Given this heterogeneity among devices, it is very challenging to develop generic endpoint security solutions for IoT devices. Although there are endpoint security solutions available for multi-purpose IoT devices, these solutions are not commonly used~\cite{isAntiVirusGood, antivirusForAndroid}. There have been reports of incidents where endpoint security solutions could not detect smartphone applications, which were involved in stealing user data or performing similar attacks~\cite{pan2018panoptispy}. 
This heterogeneity also affects the network traffic patterns of the devices, making it infeasible to collect and maintain traffic signatures databases, which are used by signature-based network security solutions~\cite{6517052, Feamster:2010:OHN:1851307.1851317}. 

Second challenge is related to the communication patterns of the IoT devices. Because IoT devices often need to interact with other devices in the same network, the IoT devices cannot be completely isolated from each other. However, simultaneously, one cannot blindly trust every device in the local network. Thus, it is not enough to simply protect the perimeter of the local network, but instead, also the internal traffic would need to be monitored.

Thirdly, traffic analysis should be performed on network gateways to address privacy and latency concerns. Typical SOHO networks are set up using low-cost network gateways with constrained hardware resources. This means that the in order to perform traffic classification on these gateways, classification algorithms should be lightweight. Furthermore, these gateways should support automated configuration because it is clear that having a dedicated administrator for every edge network is an unrealistic requirement~\cite{Feamster:2010:OHN:1851307.1851317}. 
These devices should not require much manual configuration and any required configuration changes should be easy for the users to make --- it is well known that due to poorly designed interfaces and lack of support for automated configuration, the networks are rarely configured by users~\cite{7980167, Feamster:2010:OHN:1851307.1851317, Hafeez:2016:STS:3010079.3012014}. 

Based on this discussion, the basic set of requirements for a security solution addressing the challenges posed by IoT can be summarized as follows.
\begin{itemize}
\item It should be easy to deploy and operate with minimal manual effort. Meanwhile, it should be low cost with limited resource footprint.
\item It should be able to monitor inter as well as intra-network communications to detect various (\classB and \classM) types of network traffic generated by connected devices.
\item Traffic analysis should be performed close to edge networks to immediately detect and block any attacks. Meanwhile, it should be easy to share the data needed to detect these attacks, among network gateways. 
\end{itemize}

\section{System Design}
\label{sec:sys_design}

\begin{figure}[h]
  \begin{center}
    \includegraphics[width=0.8\linewidth,bb=10 10 1250 1250]{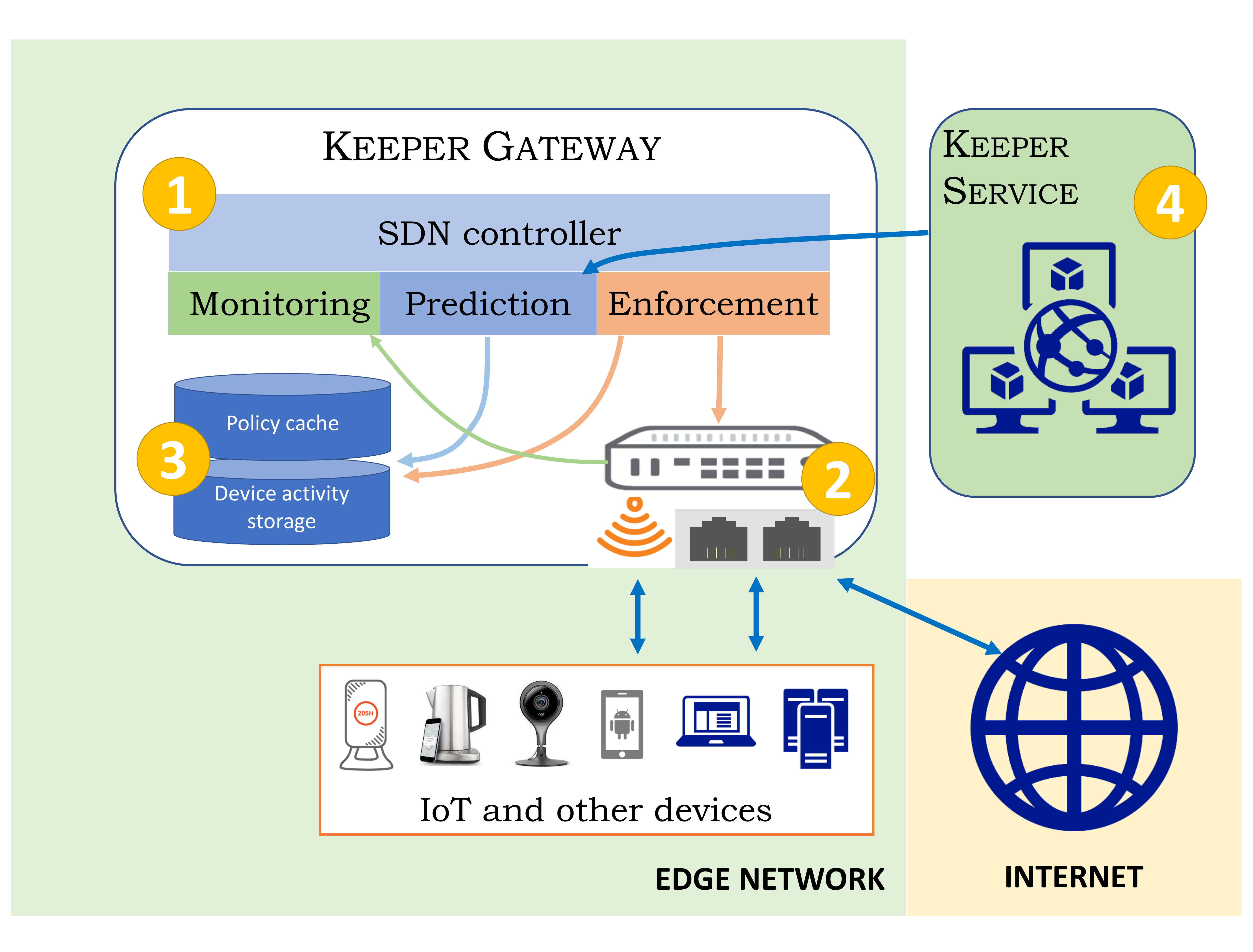}
  \end{center}
  \caption{\sysname architecture, where \frontend performs traffic monitoring and classification. Controller (1) is responsible for traffic management at OF switch (2), traffic classification, caching (3) and enforcement of \policies. \backend (4) is used for support operations}
  \label{fig:sys-overview}
\end{figure}

\sysname consists of two primary components, \frontend and \backend, as shown in Fig.~\ref{fig:sys-overview}. 
\frontend is a redesigned gateway used to set up edge networks. It also performs traffic classification and \policy enforcement for traffic filtering.
\backend is a cloud service assisting various functionalities of \frontend. 
This two--tier design achieves low cost and high scalability, to provide enterprise--grade security at only a fraction of cost. 

\subsection{\frontend}
\label{sec:design_frontend}

\frontend is a lightweight network gateway designed to be agnostic of the underlying hardware, such that it can be deployed using a \wifi access point~\cite{7980167} or single board computer, for example, Raspberry--PI (\rpi).

In principle, \frontend is a gateway used to setup edge networks. In addition to routing and switching, this gateway is responsible for traffic monitoring, anomaly detection, and policy enforcement to manage network access at per--device granularity.
Meanwhile, the initial model training, state management, and remote administration is performed with the help of \backend.

\frontend runs an SDN controller and Open vSwtich (\ovs) to perform traffic monitoring, anomaly detection and traffic filtering. 
Other than the basic routing functions, the three key modules that are operated by the SDN controller are \textit{traffic monitoring}, \textit{anomaly detection}, and \textit{\policies enforcement}. 
\textit{Monitoring} module inspects all intra and inter--network traffic flows to maintain up-to-date information about the network behavior of all devices connected to the network. 
\textit{Detection} module uses the data from monitoring module to identify if any given traffic flow is malicious. The classification model, which is used to identify different types of traffic flows, is maintained by \frontend.
\textit{Enforcement} module is responsible for setting up network access control to restrict network activity of any device exhibiting anomalous network behavior. This module uses a set of \policies to generate flow table entries and deploy them at the \ovs, to perform traffic filtering.

For every traffic flow, enforcement module looks for a relevant \policy from cache. If there are multiple matches, the most specific \policy is used for setting up flow table entries to handle the flow. Otherwise, if no relevant \policy is found, the detection module analyzes the traffic flow to identify its type. The result of analysis is cached in form of a \policy and respective flow table rules are deployed, by the enforcement module, at the switch handling given traffic flow.

In current description, both the controller and \ovs run on the same gateway. However, \sysname architecture supports deployments where a single instance of \frontend manages multiple OpenFlow--enabled switches in the network. In such cases, traffic from all the switches will be classified and managed by the controller running on \frontend.

\subsubsection*{Caching}
In general, majority of traffic in edge networks is destined to a handful of cloud services. In case of a new traffic flow, there is high probability that subsequent flows in the session are related to same network activity. The \classB network activity for single--purpose IoT devices is also fairly limited. 

By caching the \policies relevant to these frequent traffic flows, we can greatly reduce the number of traffic classification operations, thereby, reducing the latency experienced by users as well as limiting the resource consumption. The impact of caching on latency and resource consumption are discussed in detail in Sect.~\ref{sec:eval_net_perf}.

Caching can be implemented using hash table data structure, to achieve time and space complexity of $\mathcal{O}(1)$ and $\mathcal{O}(n)$ respectively. The storage consumption can further be limited by associating an \textit{expiry time} to each \policy stored in cache. This expiry time is refreshed every time \policy is used to set up filtering for some traffic flow. Once this time period expires, \policy is removed from cache. The optimal choice for expiry time depends on the underlying network traffic patterns and storage capacity available for cache.

\subsubsection*{Management API}
Unlike traditional gateways and routers, \frontend does not run a local web server for hosting gateway management portal. This design choice was made to reduce the attack surface.  Setup and configuration changes are performed using a mobile application, which communicates with the gateway using low-power protocols such as BLE. This requires physical proximity of user to make any configuration changes to the gateway. The requirement of physical proximity bars any untrusted entity from accessing the management interface over the network.
Sect.~\ref{sec:backend_design} discusses how the users can perform configuration changes, when they are connected to some other network. 

\subsection{Adhoc Overlay Networks}
\label{sec:aon_design}
It is a common requirement for gateways and routers to support multiple networks for different kinds of devices connected to the network. Although it is possible to run multiple networks on legacy gateways using VLANs or multiple SSIDs, there is only a limited number of VLANs and SSIDs\footnote{Raspberry PI supports 1 SSID using built-in WLAN interface. Commercially available routers for non-enterprise networks can support upto 8 SSIDs} supported by any router or access point available in market. It is difficult to automatically setup and manage the VLANs on router and gateways typically used to deploy edge networks. In case of multiple SSIDs, client devices need to (re)associate every time SSID configuration is updated, thereby, ruining the user experience. Therefore, it is not easy to achieve per device access control using legacy gateways.

\begin{figure}[h]
  \begin{center}
    \includegraphics[width=0.45\linewidth, bb=0 0 400 400]{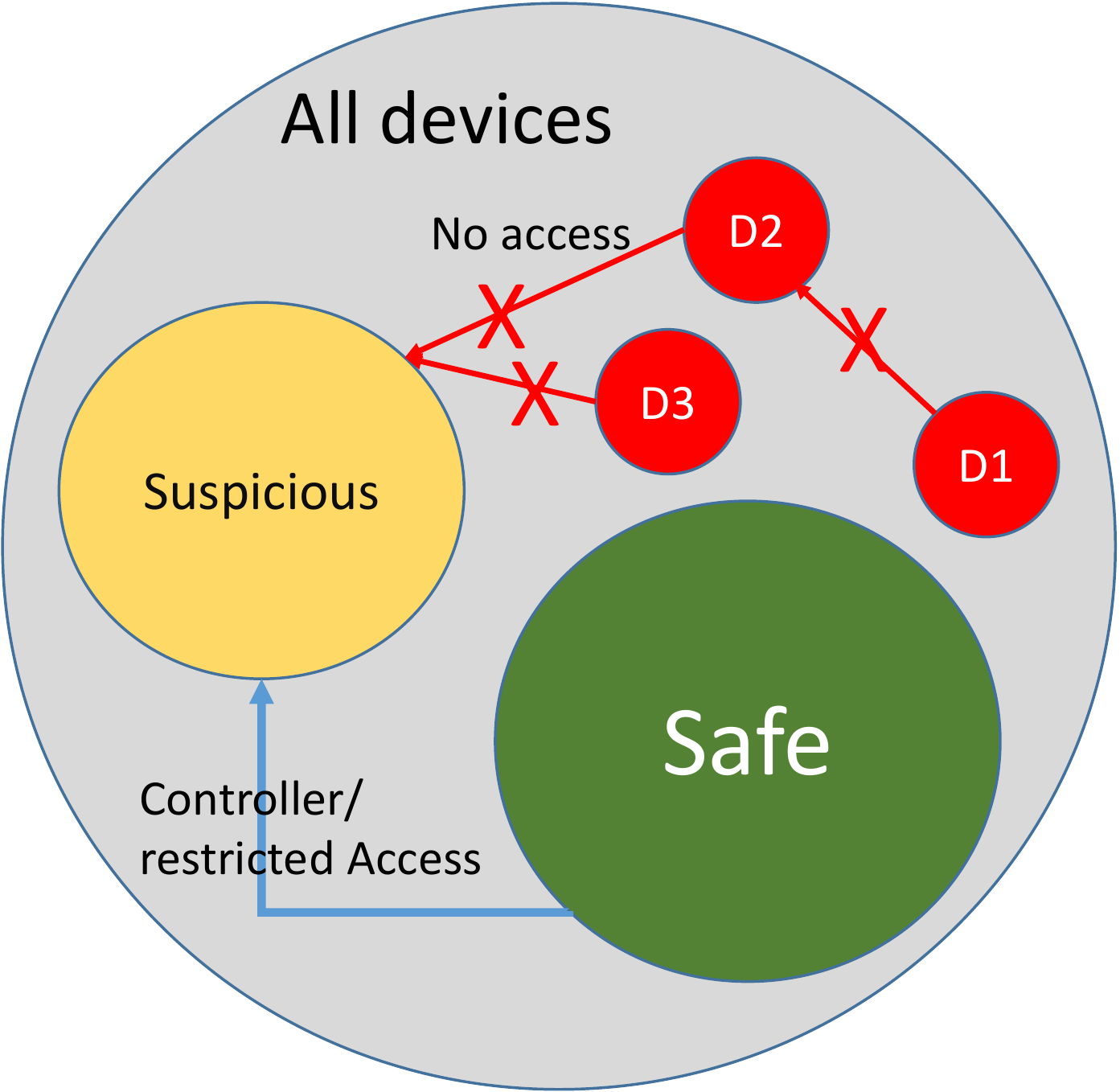}
  \end{center}
  \caption{A single SSID is partitioned in 3 AONs, where (i) D1, D2 and D3 are fully isolated and can not communicate with any other device in the same network, (ii) devices in \textit{safe} AON have full network access to Internet and other devices in same AON, but restricted access to devices in other AON. (iii) devices in \textit{suspicious} AON are partially isolated and have limited access to Internet and other devices in same AON.}
  \label{fig:aon}
\end{figure}

\sysname uses \textit{adhoc overlay networks} (AON), for creating multiple virtual networks, over a physical network.
The number of AONs is not limited by hardware as they are set up dynamically by the enforcement module. The network restrictions defined for a given AON can be updated dynamically, without requiring any action from client devices. Figure~\ref{fig:aon} demonstrates how AONs work in practice.

With AONs, it is possible to set up access control on per-device granularity. For example, a smart fridge can be restricted to communicate only with its own cloud service and owners' smartphone. AONs restrict the network access for suspicious device(s) in a way that given device retains its Internet access, but cannot attack any local or remote targets.

\subsection{\backend}
\label{sec:backend_design}
\backend is designed as a support service for \frontend operations. It does not perform traffic classification for edge networks but it hosts different services to enable seamless operations of \frontend.

The design choice of using a remote service is motivated by cost efficiency and scalability. This approach allows us to use data from multiple sources and perform various analysis to train initial classification model, used by \frontend. It also provides support for operating a multitude of services to enhance \frontend functionalities. \backend can be used to perform sophisticated analysis and operate middleboxes for re-routing suspicious traffic from edge networks, through these middleboxes. This service can be deployed within user premises or provided by a third party.

\backend is primarily responsible for maintaining up-to-date classification model for bootstrapping traffic classification on the \frontend. It also supports state management and remote administration of \frontend. 
\subsubsection*{Initial classification model}
When \frontend is set up for the first time, it recieves the initial classification model from \backend. This model is trained by \backend, using the data collected from various sources including network logs, malware databases~\cite{MalwareForResearchers}, and public vulnerability databases (CVE~\cite{cve} and CWE~\cite{cwe}). 
\frontend initially uses this model for anomaly detection, and continues to improve it using the data collected from local network traffic. 

The use of same model training technique by both \backend and \frontend improves the interoperability, in such a way that \backend and \frontend can share the trained models with each other. The trained models are shared in form of a set of rules containing feature value distributions and corresponding labels (described in Sect.~\ref{sec:prediction}). This representation is compact and thus can be easily shared among nodes without incurring significant network overhead. 

\subsubsection*{State management}
In order to support stateful recovery, \backend periodically backs up the state of \frontend, in fully encrypted form. The state information includes classification model data, \policy cache and gateway configurations. Using this data, \frontend is fully restored at any required state. Encryption and trusted platform module can be used to ensure that the state can only be restored on the same \frontend that was backed up~\cite{Dunn:2011:CMT:2028067.2028093, Ekberg:2009:EAN:2176768.2176772}. To support rapid deployment and recovery, the state can be restored or synchronized across multiple gateways simultaneously. 

\subsubsection*{Remote administration}
Since users need to be in physical proximity of \frontend to perform any configuration changes, any users connected to remote networks can perform update gateway configuration using the web-portal offered by \backend. Any configuration changes made on web-portal are installed at respective \frontend, deployed in the edge networks, by \backend. The configuration changes made through web portal are validated to make sure that they do not compromise \frontend functioning, by switching it to insecure state. The additional verification prevents any scenarios where an attacker gets access to user account or compromises the \backend, to configure all \frontend devices and disrupt their functioning. 

\subsection{Communications with user}
\label{sec:info_relay_design}
A key requirement of designing usable security solutions is finding the right balance in user experience and security. \sysname achieves this balance by maintaining minimal network access for suspicious device, so that they can continue their normal operations\footnote{Our analysis reveals that $\ge95\%$ traffic from benign IoT device is HTTP/HTTPS traffic to their cloud service, which will not be blocked, so IoT device can maintain its normal operation}, but cannot perform any attacks.

\sysname automatically detects any attacks and sets up counter measures to prevent these attacks, without user involvement. However, it is important to notify the users about blocked threats and the state of their network. For this purpose, \sysname supports two types of notifications. \textit{Passive} notifications simply inform users about the network activity, while \textit{actionable} notification require users to take an action, for example, reconfigure the device. A detailed discussion on the appropriate level of nagging and notification mechanism is outside the scope of this paper.

\section{Methodology}
\label{sec:method}
\subsection{Feature extraction}
\label{sec:feature_extraction}
{\sysname} is designed for networks with heterogeneous device base and no dedicated network security devices. It is safe to assume that there are no device logs available from most single--purpose IoT devices.
Therefore, \sysname heavily relies on feature extracted from network traffic data, to differentiate between \classB and \classM network activity.

A key challenge in feature extraction for an online traffic classification technique is to swiftly compute the statistics over incoming data (packet) stream, where packet arrival rates is very high. 
To address this, we incrementally compute a set of statistics, including number of observations $N$, sum of observations $S_{o}$ and sum of squares of observation $S_{sq}$, for all traffic streams. 
Using these statistics, we can compute mean and standard deviation for the set of observations, as shown in Eq.~\ref{eq:stats}.

\begin{equation}
\mu =  S_{o}/N
\hspace{8mm}
\sigma = \sqrt{\mid S_{sq}/N - (S_{o}/N)^2 \mid}
\label{eq:stats}
\end{equation}

\begin{table*}[th]
\caption{List of attributes extracted from network and device metadata} 
\centering
\label{tbl:features}
\begin{tabular}{p{0.18\linewidth}p{0.70\linewidth}}
\toprule
\multicolumn{1}{c}{\textbf{Type}}         & \multicolumn{1}{c}{\textbf{Feature}} \\  \midrule
Source, Destination 		& [Total, Unique] destination IP addresses \\ 
Connection counters	& [Total, Unique] source ports, destination ports, connections, (same source,\ same destination,\ same service) connections, connection durations (binned) \\ 
Packet counters		& ARP, LLC, IP(v6), ICMP(v6), EAPoL, TCP(v6), UDP(v6), HTTP, FTP, HTTPS, DHCP, (M)DNS, NTP, Router Alert, (SYN, REJ) (errors), Urgent, Padding \\ 
Data (binned)		& Total data, source to destination (SRC2DST) data, destination to source (DST2SRC), packet size  \\ 
Authentication		& [Successful, Failed] login attempts to [SSH, Service, Device] \\ \bottomrule
\end{tabular}
\end{table*}

Table~\ref{tbl:features} lists the 44 attributes obtained from network metadata and device logs. We calculate the three aforementioned statistics for all 44 attributes to get the final feature vector $\vec{F}$, where $ \vec{F}\;\epsilon\;\mathbb{R}^n \wedge n\le132$. The statistics can be summarized on per device basis using source and destination MAC, IP and ports. 
These statistics  allow us to calculate the divergence of devices' behavior, over a specific time window, from its baseline (\classB) behavior, to detect any anomalies.

In contrast to existing techniques~\cite{KDDCup99Dataset}, which use time-based aggregation to aggregate same host, same service features, we use connection based aggregation. The key limitation of time-based aggregation is that it falls short in detecting attacks with a wait mechanism, where a random delays are added in-between successive connection attempts. In comparison, connection-based aggregation aggregating the features over $n$ latest connections allows us to accommodate any random delays between successive connections. 

Table~\ref{tbl:features} lists six attributes retrieved from device logs. These attributes contain information about any login attempts, SSH connections and service discovery requests.
Such information can be useful to identify the sub-type of malicious traffic flows identified by anomaly detection technique. 
In order to correlate the data from device logs with network traffic data, time must be synchronized across the network. Time synchronization among all connected devices can be achieved using protocols such as NTP. In case no such service is running in the network, time difference between network and device logs can be manually resolved. 

\subsection{Feature Analysis}
\label{sec:feature_analysis}
We study the variance and modality of each feature to identify its contribution to anomaly detection model. Any features that do not contribute significantly to the clustering and classification process are pruned off. 
Reducing the number of features helps in speeding up the classification process and reducing the resource footprint of clustering and anomaly detection scheme. 

\begin{figure*}[th]
\begin{minipage}{.25\linewidth}
\centering
\subfloat[]{\label{fig:cdf_dest_ip}\includegraphics[scale=.23]{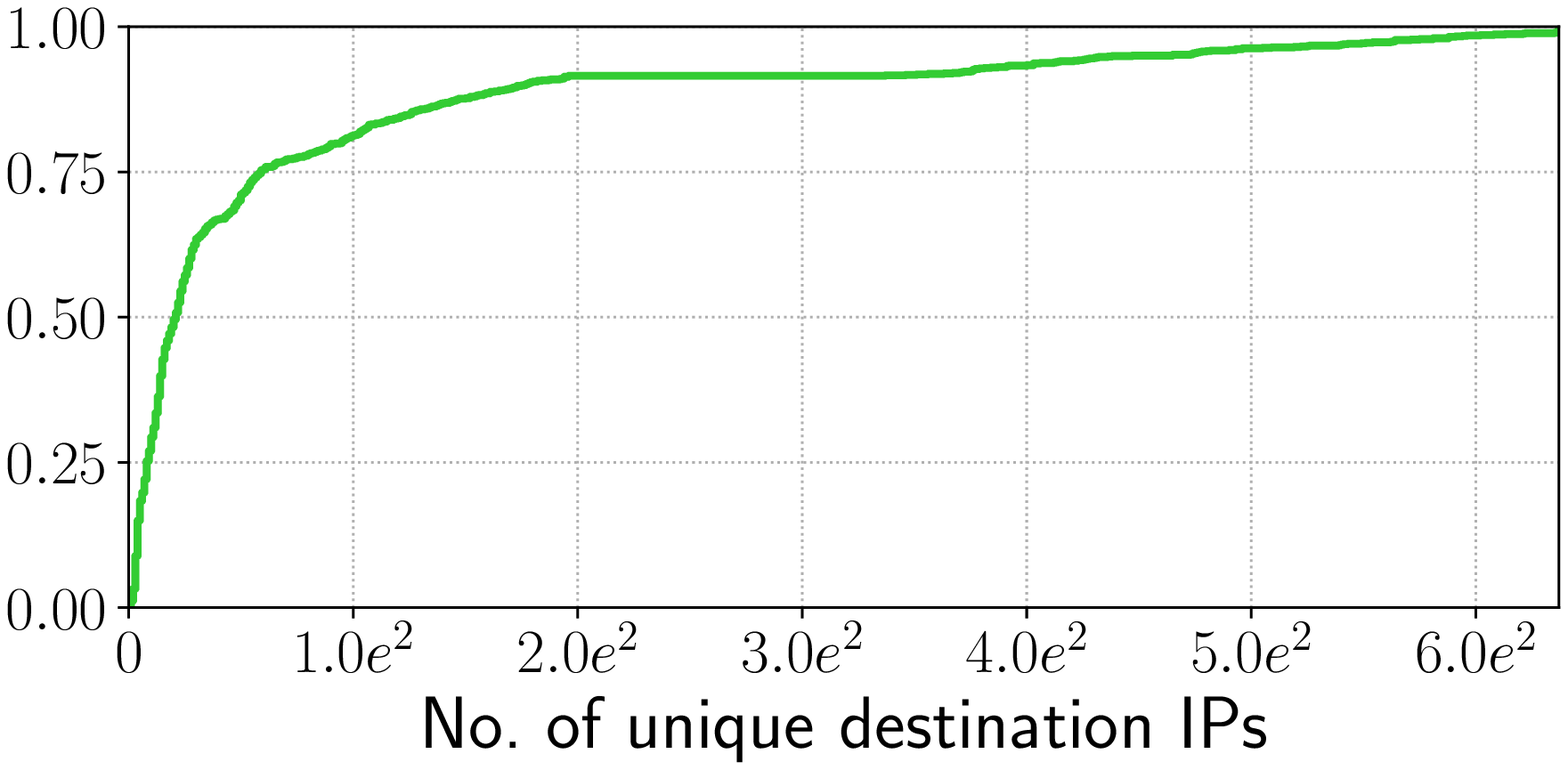}}
\end{minipage}%
\begin{minipage}{.25\linewidth}
\centering
\subfloat[]{\label{fig:cdf_dest_port}\includegraphics[scale=.23]{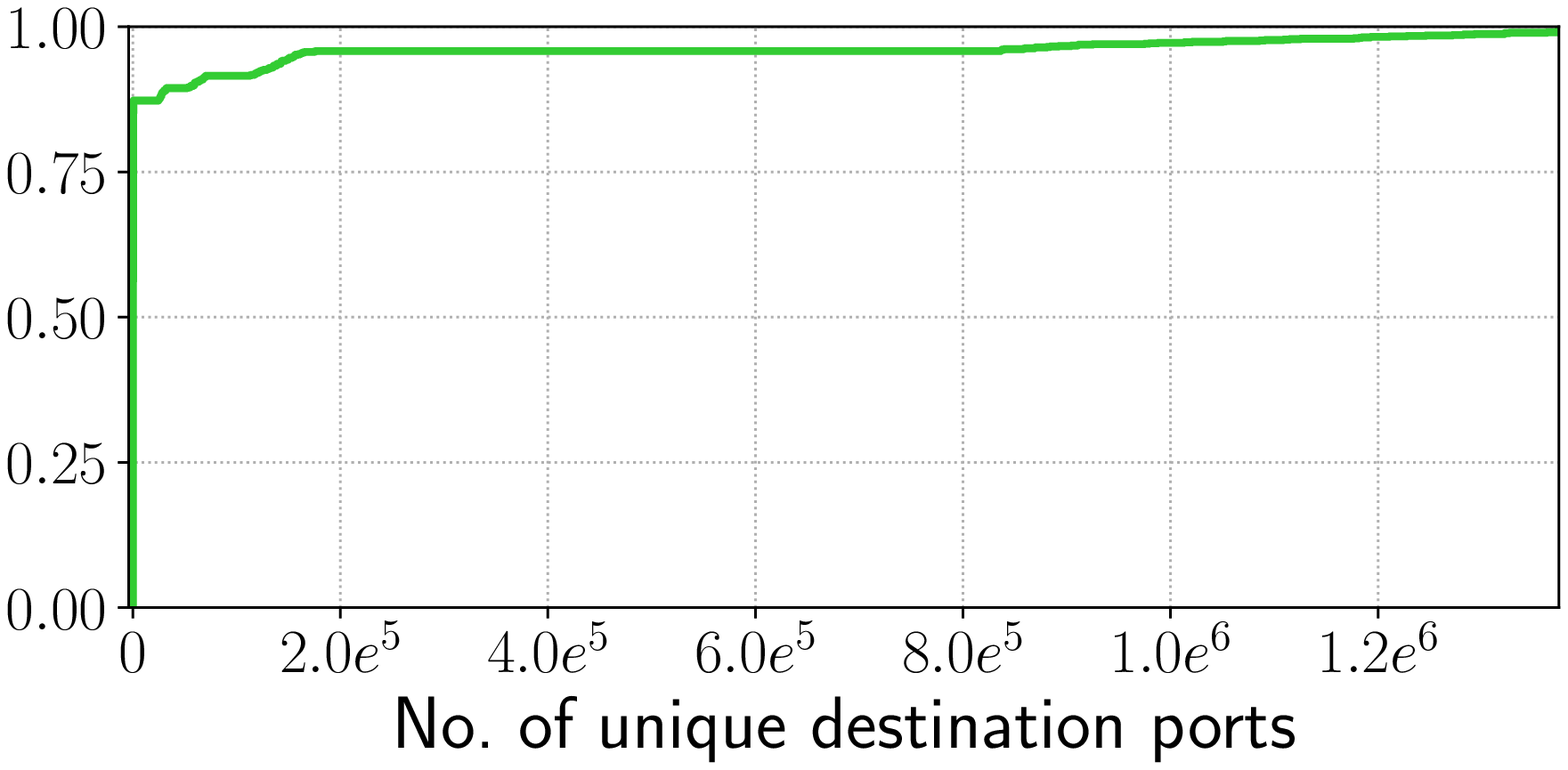}}
\end{minipage}%
\begin{minipage}{.25\linewidth}
\centering
\subfloat[]{\label{fig:cdf_conn}\includegraphics[scale=.23]{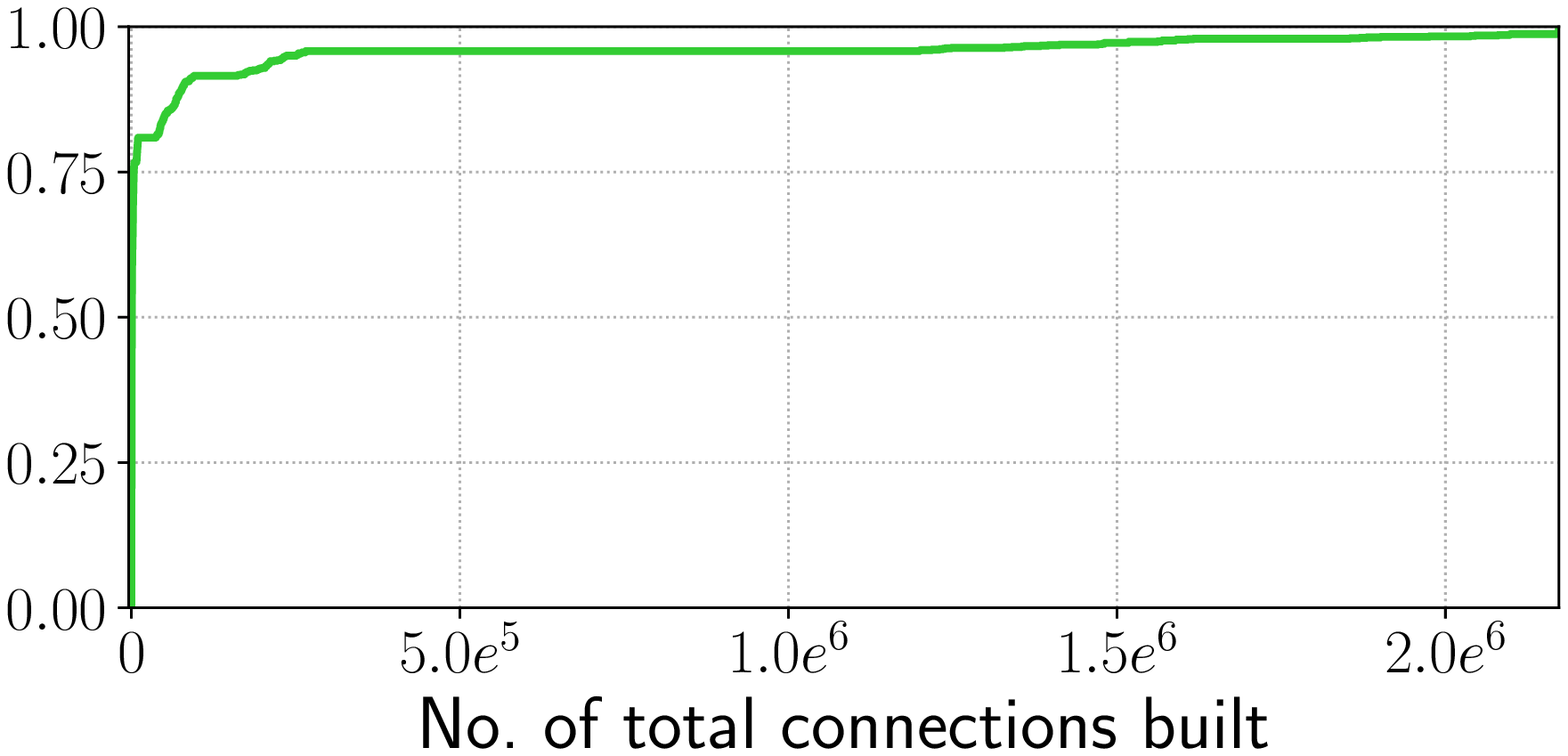}}
\end{minipage}%
\begin{minipage}{.25\linewidth}
\centering
\subfloat[]{\label{fig:cdf_ssh}\includegraphics[scale=.23]{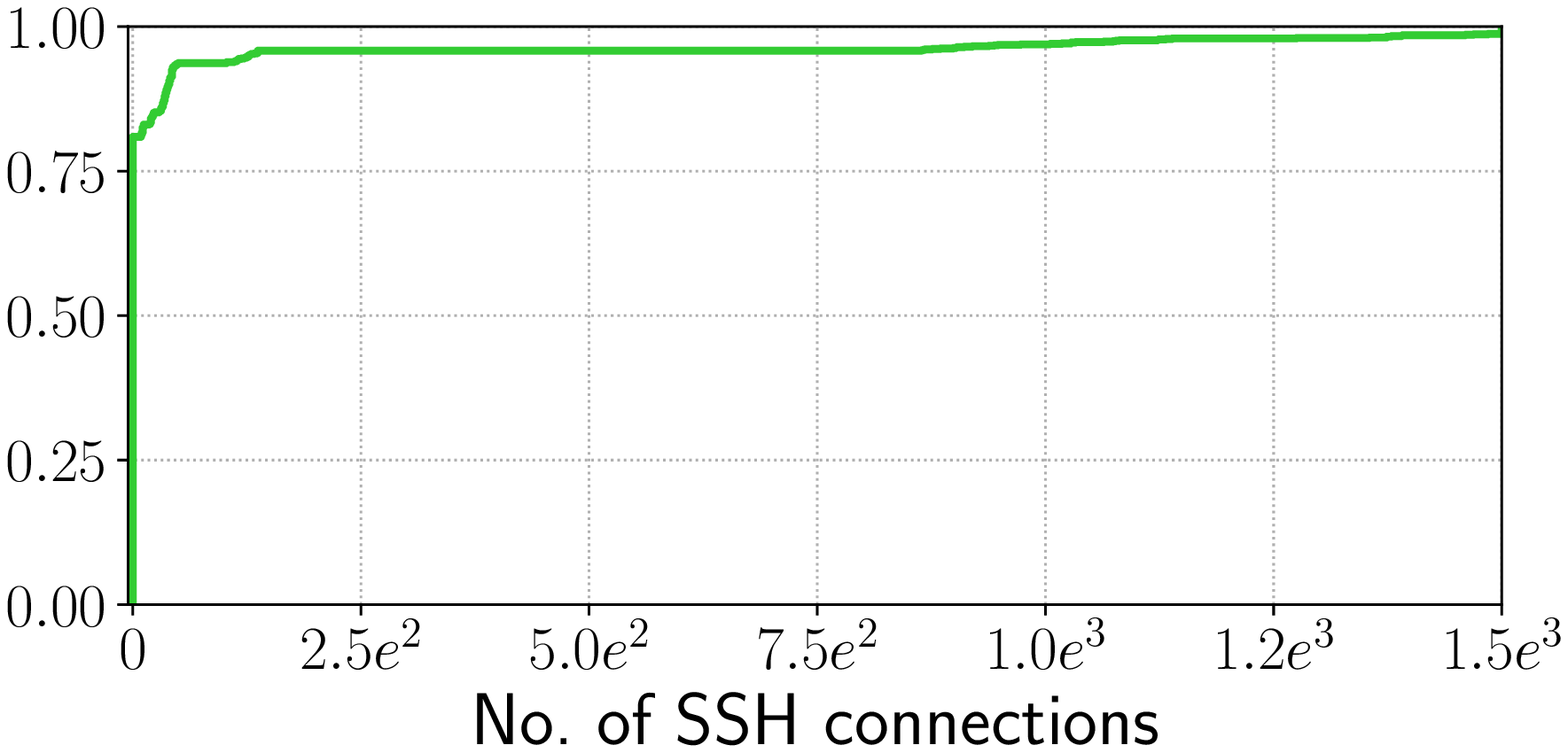}}
\end{minipage}%
\caption{CDF plots for a small subset of features corresponding to network behavior of user devices. These distributions are observed in the dataset used for training and evaluation purposes.}
\label{fig:cdf_features}
\end{figure*}

Initially, we plot cumulative distribution function (CDF) for each feature, to study its variance. Figure~\ref{fig:cdf_features} shows CDF plots for a few of the features corresponding to connections made by any device. For example, Fig.~\ref{fig:cdf_dest_ip} shows the distribution for number of unique destination IPs contacted by devices in the network. It can be observed that this distribution is not Gaussian but heavy tailed with majority of probability mass lying in smaller values. This distribution reveals that more than $70\%$ of the devices connect to fewer than $20$ unique destination IPs. On the other hand, tail of distribution contains data points where a single device may connect to more than $500$ unique destinations. Similary, Fig.~\ref{fig:cdf_ssh} shows that more than $75\%$ of the devices do not generate any SSH traffic. However, there are some devices generating a large volume of SSH traffic, indicating presence of suspicious activity in the network.  

The data points in the tail of distributions are of primary importance for anomaly detection since they capture anomalous behaviors. It is important to capture this information as it helps in differentiating anomalies from benign network behavior, during clustering.

When we jointly study the distribution of different features, it reveals interesting details about semantics of various attacks. For example, during a \textit{fuzzing attack}, when an attacker tries to login to open services on target node with brute-force, the attack is reflected in network traffic with a large number of connections between same source and destination nodes. Meanwhile, device logs from target node show a large number of failed login attempts, proving the hypothesis that an attack is undergoing against target node. 

Further analysis revealed that during such brute-force attacks, if the few initial login attempts fail, there is a high probability that attack will not succeed. This observation shows that users who change the factory default passwords choose reasonably unique passwords, which are not easy to crack using traditional dictionary or brute-force attacks. However, this claim cannot be made with absolute certainty because of limited scope of dataset.

The study of feature value distributions also reveals possible correlations among different features. For example, as the attacker scans a target node, both total number of connections initiated by the attacker node and number of connections between source (attacker) and destination (target) node increases. These correlations help us to identify and remove features containing redundant information.

\subsection{Feature reduction}
\label{sec:feature-reduction}
We use \textit{correlation-based feature selection} (CFS) and \textit{deviation method} to identify and remove any features which contain redundant information and do not contribute significantly to the anomaly detection scheme.

\begin{figure}[th]
  \begin{center}
    \includegraphics[width=0.80\linewidth]{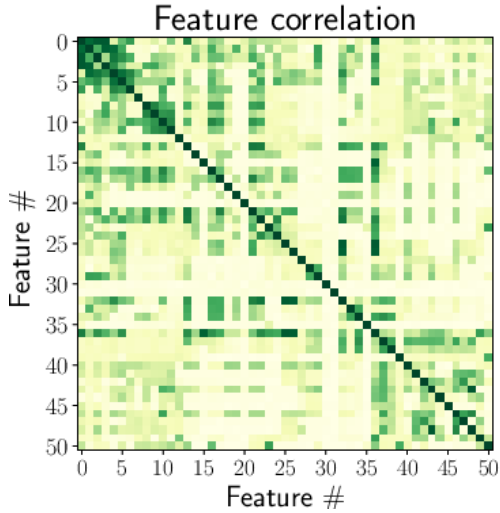}
  \end{center}
  \caption{Correlation map plot depicting linear dependence among a subset of features from $\vec{F}$}
  \label{fig:feature_corr}
\end{figure}

CFS identifies strongly correlated features by measuring their linear dependencies. The dependencies are calculated using Pearson correlation coefficient $R$ because it provides fairly accurate results with bounded feature value ranges for datasets of fairly large size. Based on the value of $R$, CFS discards one of any two features which are strongly co-related, since such two features contain redundant information and keeping both features do not offer value for anomaly detection. Figure~\ref{fig:feature_corr} shows that majority of features in our feature set $\vec{F}$ are linearly independent. However, some of the features, such as $f_{1}$--$f_{6}$, may contain redundant information and can be removed from $\vec{F}$.

Using \textit{deviation method}, 1-length frequent items sets are mined from the feature set to obtain $F_{i}=[V_{1}, V_{2}, \hdots, V_{j}]$, where $V_{j}=[f_{i}; 1\le i \le n], supp(f_i) \ge m$, $m$ is minimum support and $f_{i}$ is frequent item. The deviation range for each feature is calculated as $D_{V_{j}} = [f_{max}, f_{min}]$, where $f_{max} = \max(f_{j})$ and $f_{min} = \min (f_{j})$, for all \classB and \classM traffic classes. If the deviation range of a feature is similar for all classes, the feature is considered non-contributing feature and removed from the feature set.

In order to make sure that no feature over-influences clustering, all feature values are normalized to range $[0,1]$. After clustering, normalized feature score are computed, for each feature, in all clusters. Any features, such as REJ errors, with same scores (within a defined tolerance) in multiple clusters are considered non-contributing feature and removed from the final feature set.

\subsection{Clustering}
\label{sec:clustering}
We use fuzzy C-mean (FCM) clustering algorithm to partition the data points based on their mutual likeliness. 
During clustering, initially a membership value is assigned to all data points $X_{j}\;(j=1, 2, ..., n)$, for all clusters $C_{i}\;(i=1,2,...,c)$. Each data point $X_{j}$ is represented as $\left(f_{j}^{(1)}, f_{j}^{(2)},..., f_{j}^{(k)}, ..., f_{j}^{(h)}\right)$ where $f_{j}^{(k)}$ is value for $k^{th}$ feature in $X_{j}$ and $1\leq k \leq n,\quad n=len(\vec{F})$.

The membership value for $X_{j} \in C_{i}$ is given as $\mu_{ij}$, where $0 \leq \mu_{ij} \leq 1, \sum_{i=1}^{c} \mu_{ij}=1 \quad \forall \ 1\le i \le c \wedge 1\le j \le n$. The membership value $\mu_{ij}$ for each data points and cluster centers $V_{i}$ for each cluster are optimized using Eq.~\ref{eq:membership-ftn} and Eq.~\ref{eq:cluster-centers} respectively, in order to minimize objective function given in Eq.~\ref{eq:obj-ftn}.
 
\begin{align}
\mathsmaller{
\mu_{ij} =
\left(
\sum\limits_{d=1}^{c}
     \left(\frac{\norm{V_{i}-X_{j}}}{\norm{V_{d}-X_{j}}}\right)^{\mathsmaller{2}\mathsmaller{m-1}}
\right)^{-1}
}
\hspace{0mm} & , \mathsmaller{\genfrac{}{}{0pt}{}{1 \leq i \leq c}{1 \leq j \leq n}}
\label{eq:membership-ftn}\\
\mathsmaller{
	V_{i} = \frac{(\sum\limits_{j=1}^{n}(\mu_{ij})^{m}\times X_{j})}{\sum\limits_{j=1}^{n}(\mu_{ij})^{m}}
}
\hspace{12mm} & , \mathsmaller{\genfrac{}{}{0pt}{}{1 \leq i \leq c}{1 \leq j \leq n}}
\label{eq:cluster-centers}
\end{align}

\begin{equation}
J_{m}=\sum_{i=1}^{c}\sum_{j=1}^{n}\mathlarger{\mu}_{ij}^{m}\norm{V_{i}-X_{j}}^2
\label{eq:obj-ftn}
\end{equation}
where $m$ is fuzziness index~\cite{Zhou2014} and $\norm{V_{i}-X_{j}}$ is the Euclidean distance between cluster center $V_{i}$ (for cluster $C_{i}$) and data point $X_{j}$.

A label is assigned to each cluster based on normalized feature scores observed in the given cluster. These labels correspond to different types of benign and malicious network traffic. Each cluster can be represented as a rule, where feature scores represent antecedent variables and cluster label is the consequent variable. The set of rules, obtained as output of clustering process, is used by FIS to perform anomaly detection. 

\subsection{Parameter Selection}
\label{sec:param-selection}
\begin{figure*}[th]
\begin{minipage}{.25\linewidth}
\centering
\subfloat[]{\label{fig:nbclust}\includegraphics[scale=.27]{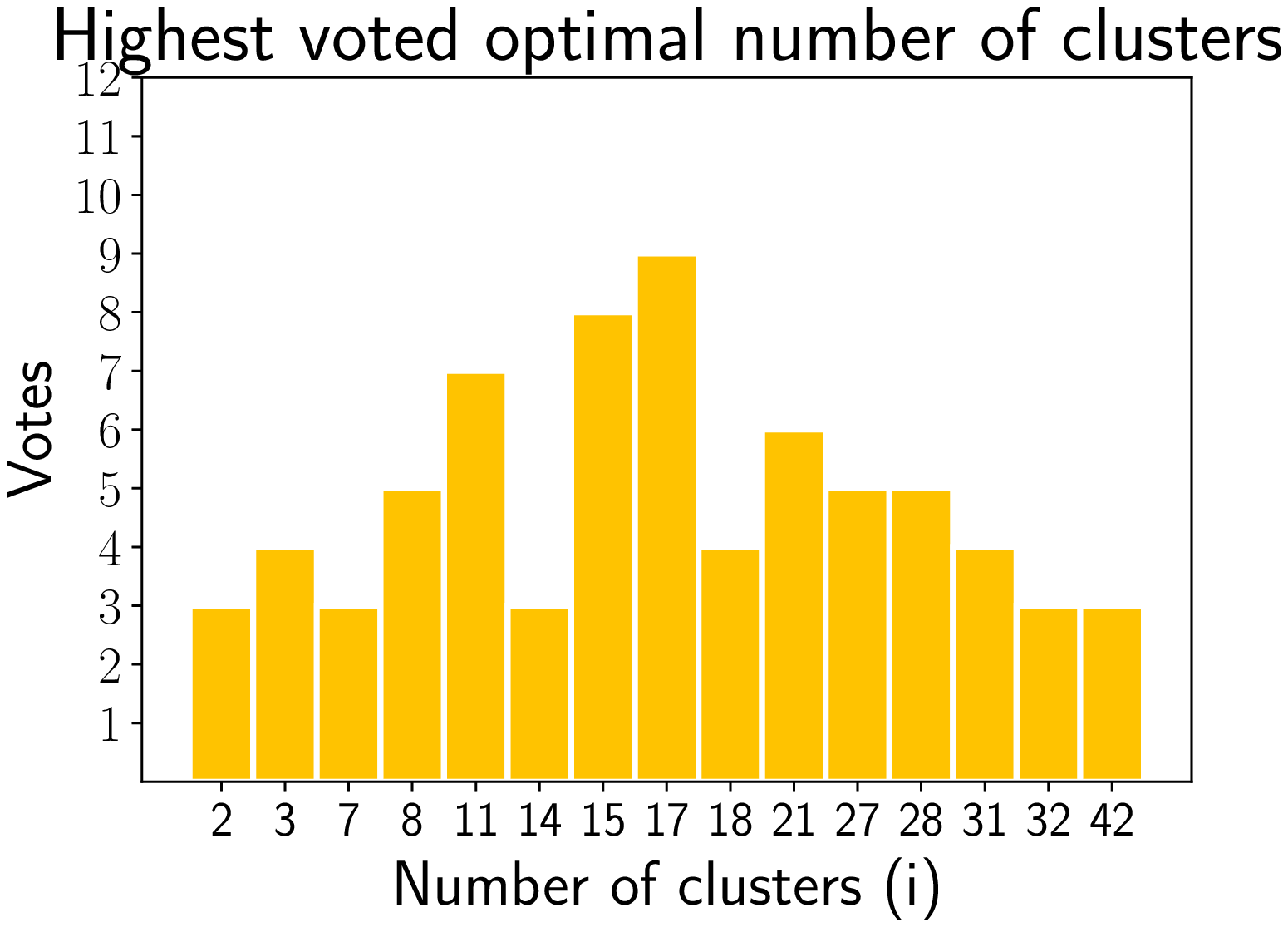}}
\end{minipage}%
\begin{minipage}{.25\linewidth}
\centering
\subfloat[]{\label{fig:elbow}\includegraphics[scale=.27]{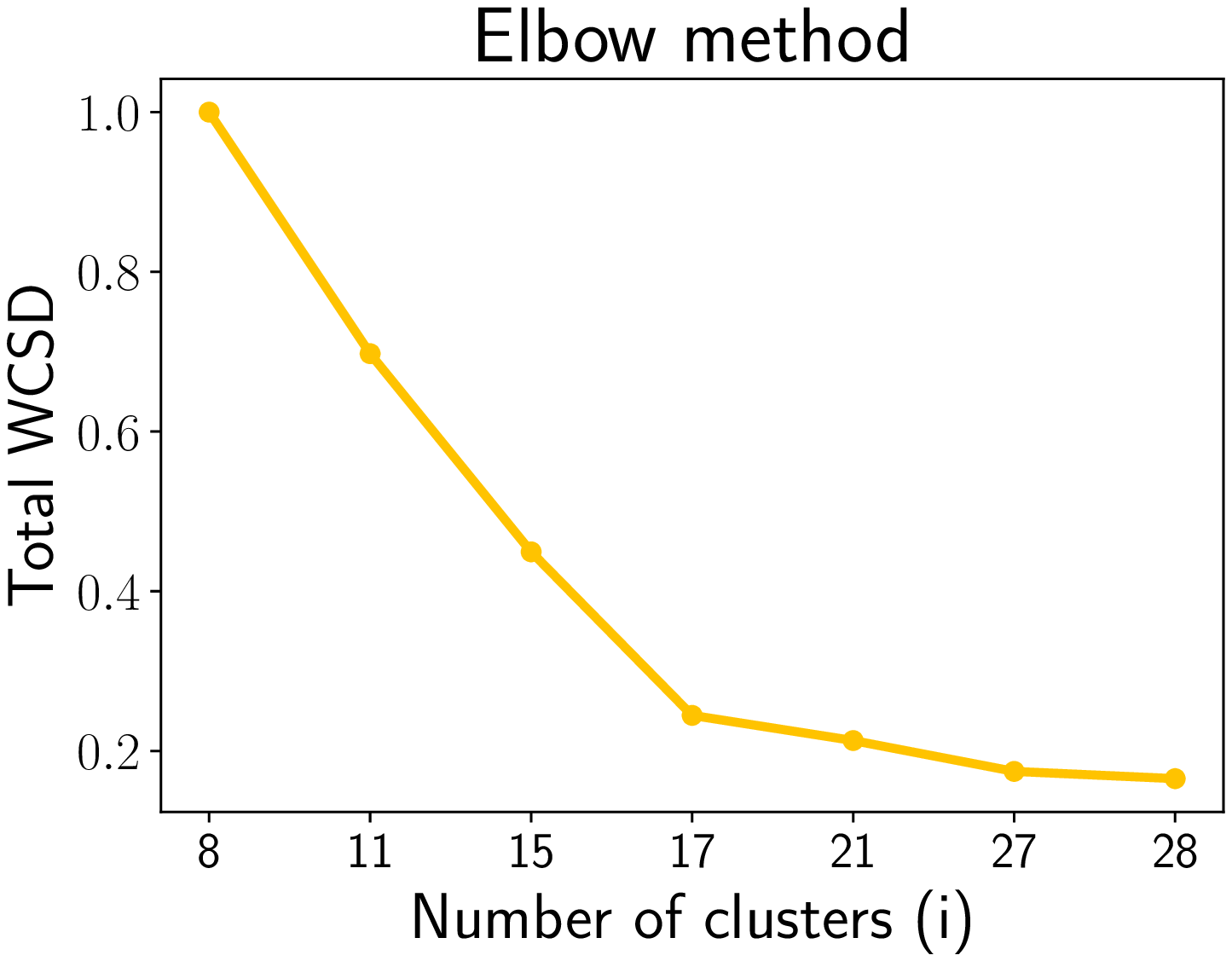}}
\end{minipage}%
\begin{minipage}{.25\linewidth}
\centering
\subfloat[]{\label{fig:sillhouette}\includegraphics[scale=.27]{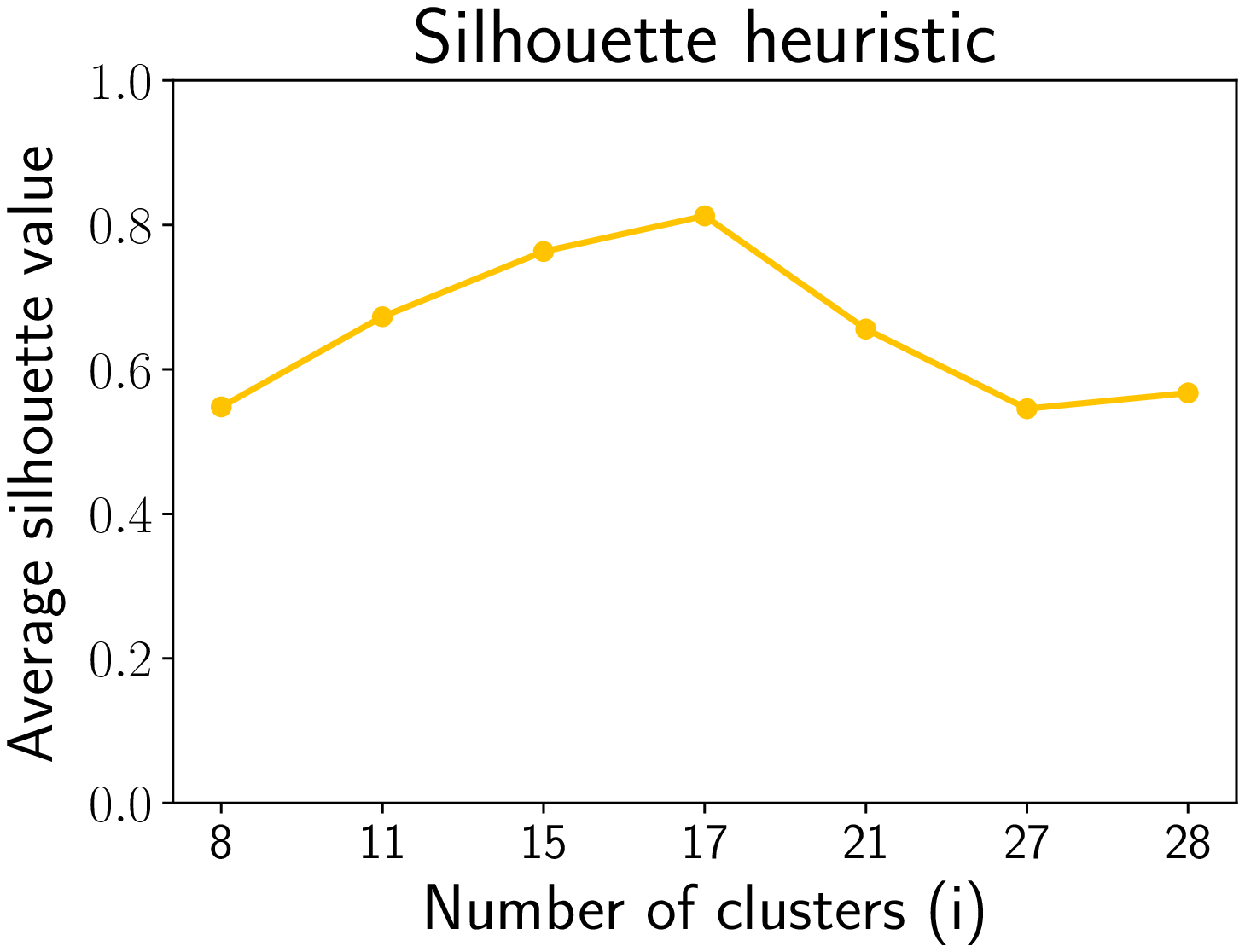}} 
\end{minipage}%
\begin{minipage}{.25\linewidth}
\centering
\subfloat[]{\label{fig:gaps}\includegraphics[scale=.27]{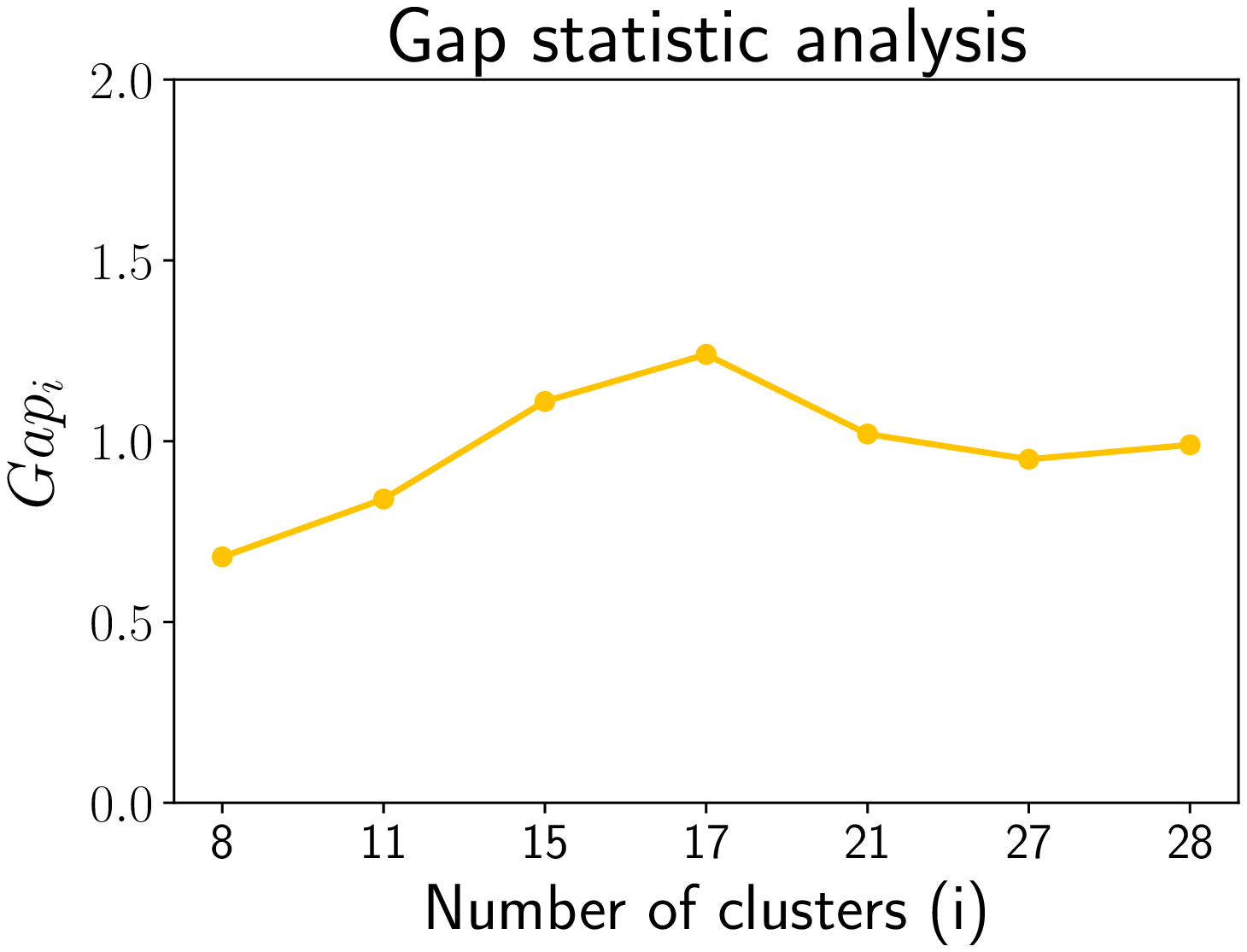}}
\end{minipage}%

\caption{(a) Candidate values for optimal number of clusters ($i$), based on voting results (minimum 3 votes), obtained using \texttt{NbClust} package. Elbow method (b), average silhouette heuristics (c) and gap statistics (d), were used to identify the optimal value of $i$, out of top 8 candidate values for $i$ shown in (a).}
\label{fig:params}
\end{figure*}

The choice of number of clusters $i$ can affect the performance of anomaly detection technique. Therefore, we use both direct and statistical testing methods to choose the optimal value of $i$.
Initially, we compute 30 different indices for a range of possible values for $i$, using \texttt{NbClust} package~\cite{JSSv061i06}. Our implementation uses \textit{agglomeration} method for cluster analysis using Wards' linkage method and \textit{euclidean} distance metric. 
Figure~\ref{fig:nbclust} shows the number of votes (minimum 3 votes) received by each possible choices for optimal number of clusters. One vote represents that one of the 30 indices suggests that the given value of $i$ is the optimal number of clusters. A detailed discussion on various indices computed by \texttt{NbClust} package is out of scope for this paper.

Based on the voting results of \texttt{NbClust}, we select the top eight candidate values of $i$ and analyze them using \textit{elbow method} and \textit{average silhouette heuristic}~\cite{ROUSSEEUW198753}. These two methods provide a measure of global clustering characteristic.
For \textit{elbow} method, within-cluster-sum-of-distances (WCSD) is calculated using Eq.~\ref{eq:wcsd}, where $c$ is the number of clusters, $S_{i}$ is the set of data points belonging to $i^{th}$ cluster, and $x_{ki}$ is the $k^{th}$ variable of $V_{i}$. 

\begin{equation}
\textit{WCSD} = \sum\limits_{i=1}^{c}\sum\limits_{j \in S_{i}}\sum\limits_{k=1}^{p} \norm{x_{ki} - x_{ji}}
\label{eq:wcsd}
\end{equation}  

Silhouette heuristics are calculated using Eq.~\ref{eq:silhouette_value}, where ${a(x)= \frac{1}{k}\sum_{j=1}^{k} \norm{x-p_{j}}}$, $p_{j} \in C_{i} \wedge x \in C_{i}$. Similarly, ${b(x) = \frac{1}{k}\sum_{j=1}^{k}\norm{p_{k} - x}}$, where $p_{k} \in C'_{i}$ and $C'_{i}$ is the closest neighboring cluster for $x$ such that $C'_{i} = C_{i} \in C$ with $\min(\norm{x - V_{i}})\; \forall \; C_{i} \in C\wedge x \not\in C_{i}$. Figure~\ref{fig:params} shows that both \textit{elbow} and \textit{silhouette} method suggest $i=17$ as optimal value for $i$. 

\begin{equation}
s(x) = \frac{(b(x)- a(x))}{\max\big(a(x),b(x)\big)}
\label{eq:silhouette_value}
\end{equation}

We also studied \textit{gap statistic method}~\cite{GapAnalysis} to get a statistical formulation of WCSD and silhouette statistics. In general, the optimal value for $i$ should maximize the gap statistic as well as silhouette values, while minimizing WCSD. Using 1-standard-error method ~\cite{GapAnalysis}, gap statistics analysis suggests $i=17$ as optimal number of clusters for given scenario. 

\subsection{Anomaly detection}
\label{sec:prediction}
\sysname uses \textit{Fuzzy interpolation scheme}~\cite{4358815, 5346229} (FIS) to identify the type of given traffic flow in the network. FIS uses the sparse fuzzy rule base consisting of $n$ rules $\left(n=c\right)$, obtained from clustering, to identify the type of traffic flows in the network.
The set of rules is represented as.

$\mathsmaller{Rule\;1:\; if\;f_{1} \in A_{11}, f_{2} \in A_{21}, \; ... \;, f_{k} \in A_{k1}, \; ... \;, f_{h} \in A_{h1}\;\Longrightarrow\;y\;\in\;O_{1}}$

$\mathsmaller{Rule\;2:\; if\;f_{1} \in A_{12}, f_{2} \in A_{22}, \; ... \;, f_{k} \in A_{k2}, \; ... \;, f_{h} \in A_{h2}\;\Longrightarrow\;y\;\in\;O_{2}}$

\vspace{-0.5mm}
\hspace{0.20\textwidth}\vdots

\vspace{-1.5mm}
$\mathsmaller{Rule\;Q:\; if\;f_{1} \in A_{1q}, f_{2} \in A_{2q}, \; ... \;, f_{k} \in A_{kq}, \; ... \;, f_{h} \in A_{hq}\;\Longrightarrow\;y\;\in\;O_{q}}$

$\mathsmaller{Observation:\;f_{1} \in A_{1}^{*},\; f_{2}\in A_{2}^{*}, \; ... \;, f_{k}\in A_{k}^{*}, \; ... \;, f_{h} \in A_{h}^{*}}$

\vspace{-2mm}\rule{8cm}{0.4pt}

\vspace{-0.5mm}
$\mathsmaller{Conclusion:\;y = O^{*}}$

\noindent
where $R_{i}$ ($1 \leq i \leq Q$) is $i^{th}$ rule generated from cluster $C_{i}$. 
$A_{ki}$ and $O_{i}$ are triangular fuzzy sets for $k^{th}$ antecedent feature $f_{k}, 1 \leq k \leq h$ and consequent variable $y$ respectively.
For any new observation, $A_{k}^{*}$ and $O^{*}$ are triangular fuzzy sets for antecedent and consequent variable obtained as a result of interpolation of spare fuzzy rule base. 

A fuzzy triangular set $A$ is represented using three characteristic points $a$, $b$, and $c$, where $b$ is \textit{center point} with maximum membership value and $a$, $c$ are \textit{left}, \textit{right} points respectively, with minimum membership value.
The characteristic points $a_{ki}$, $b_{ki}$, $c_{ki}$ for fuzzy set $A_{ki}$ of $k^{th}$ antecedent feature $f_{k}$ in rule $R_{i}$ are calculated as:

\begin{equation}
\mathsmaller{
    b_{ki}=f_{q}^{(k)}
}, \quad where\;\mu_{iq} = \max_{1 \leq j \leq n}  \mu_{ij},
\label{eq:bkf_for_Aki}
\end{equation}

\begin{equation}
\mathsmaller{
    a_{ki}=\dfrac{\mathlarger{\sum}\limits_{j=1,2,...,n\;and\;f_{j}^{(k)} \leq b_{ki}} \mu_{ij} \times f_{j}^{(k)}}{\mathlarger{\sum}\limits_{j=1,2,...,n\;and\;f_{j}^{(k)} \leq b_{ki}} \mu_{ij}}
},
\label{eq:akf_for_Aki}
\end{equation}

\begin{equation}
\mathsmaller{
    c_{ki}=\dfrac{\mathlarger{\sum}\limits_{j=1,2,...,n\;and\;f_{j}^{(k)} \geq b_{ki}} \mu_{ij} \times f_{j}^{(k)}}{\mathlarger{\sum}\limits_{j=1,2,...,n\;and\;f_{j}^{(k)} \geq b_{ki}} \mu_{ij}}
} ,
\label{eq:ckf_for_Aki}
\end{equation}

where $b_{ki}$ has membership value of 1 and $a_{ki}$ and $c_{ki}$ have membership value of 0. $f_{j}^{(k)}$ is the $k^{th}$ feature's value in sample $X_{j}$ with $1 \leq k \leq h$. The defuzzified value of a triangular set $A$ is calculated as  

\begin{equation}
D_{f}(A)=\dfrac{(a + 2\times b + c)}{4}
\label{eq:defuzz_value}
\end{equation}

Similarly, the characteristic variable $a_{i}, b_{i}, c_{i}$ for consequent variable $B_{i}$ for $R_{i}$ are calculated as:
\begin{equation}
\mathsmaller{b_{i}=O_{q}}, \quad where\;\mu_{iq} = \max_{1 \leq j \leq n}  \mu_{ij} ,
\label{eq:bf_for_Bi}
\end{equation}

\begin{equation}
\mathsmaller{
    a_{i}=\dfrac{\mathlarger{\sum}\limits_{j=1,2,...,n\;and\;O_{j} \leq b_{i}} \mu_{ij} \times O_{j}}{\mathlarger{\sum}\limits_{j=1,2,...,n\;and\;O_{j} \leq b_{i}} \mu_{ij}}
} ,
\label{eq:af_for_Bi}
\end{equation}

\begin{equation}
\mathsmaller{
    c_{i}=\dfrac{\mathlarger{\sum}\limits_{j=1,2,...,n\;and\; O_{j} \geq b_{i}} \mu_{ij} \times O_{j}}{\mathlarger{\sum}\limits_{j=1,2,...,n\;and\;O_{j} \geq b_{i}} \mu_{ij}}
} ,
\label{eq:cf_for_Bi}
\end{equation}

where $O_{j}$ is expected output class for $X_{j}$ and $1 \leq i \leq c$. 

The membership value for input feature $f_{j}^{(k)}$ is $\mu_{A_{k,i}} (f_{j}^{(k)})$, where ${\min\limits_{1 \leq k \leq h} \mu_{A_{k, i}} (f_{j}^{(k)}) > 0}$,~${1 \leq i \leq p}$,~and $p$ is the number of activated fuzzy rules. The inferred output $O_{j}^{*}$ based on fuzzy rules activated by $f_{j}^{(1)}, f_{j}^{(2)}, ..., f_{j}^{(h)}\; \in\; X_{j}$ is calculated as,

\begin{equation}
\mathsmaller{
    O_{j}^{*} = \dfrac{\mathlarger{\sum}\limits_{i=1}^{p} \min\limits_{1 \leq k \leq h} \mathlarger{\mu}_{A_{k,i}} (f_{j}^{(k)}) \times D_{f}(B_{i})}{\mathlarger{\sum}\limits_{i=1}^{p}\min\limits_{1 \leq k \leq h} \mathlarger{\mu}_{A_{k,i}} (f_{j}^{(k)})}
}
\label{eq:calc_oj}
\end{equation}

$D_{f}(B_{i})$ is defuzzified value for consequent fuzzy set, in $R_{i}$ activated by $X_{j}$ inputs and it can be calculated using Eq.~\ref{eq:defuzz_value}.
We calculate the weight $W_{i}$ of activated rule $R_{i}$, such that $0 \leq W_{i} \leq 1$,
$\sum\limits_{i=1}^{c}W_{i} = 1$, on the basis of input observations ${x_{1}=f_{j}^{(1)}, x_{2}=f_{j}^{2)}, ..., x_{h}=f_{j}^{(h)}}$ as: 

\begin{equation}
\mathsmaller{
    W_{i} = \left(\mathlarger{\sum}\limits_{d=1}^{c} \left(\dfrac{\norm{r^{*} - r_{i}}}{\norm{r^{*} - r_{d}}} \right)^{2} \right)^{-1}
},
\label{eq:calc_weight}
\end{equation}
where $r^{*}$ is the input feature vector $\left(f_{j}^{(1)}, f_{j}^{(2)}, ..., f_{j}^{(h)}\right)$ and $r_{i}$ is set of defuzzified values of $A_{ki}$ in $R_{i}$.
$\left(D_{f}\left(A_{1, i}\right), D_{f}\left(A_{2, i}\right), ..., D_{f}\left(A_{h, i}\right)\right)$,  $1\leq k \leq h$.
 
The final inferred output is calculated as
\begin{equation}
O^{*}_{j} = \sum\limits_{i=1}^{c} W_{i} \times D_{f}\left(B_{i}\right)
\label{eq:final_oj}
\end{equation}

\section{Dataset}
\label{sec:dataset}
We have deployed a real world testbed for data collection and system evaluation. The testbed consists of more than 40 consumer IoT devices including single-purpose and multiple-purpose devices. All devices mainly use wireless mode for network connectivity, while some devices also support wired connectivity as well. Some of the devices use BLE for device pairing purposes only. Other low-energy communication protocols such as Weave, Zigbee are mainly used by devices to communicate with IoT hubs. 
The set of common vulnerabilities found in these devices include factory-default, commonly-used login credentials, open, unfiltered ports, and terminal access without password.

The choice of multi-purpose devices such as phones and PCs, in the testbed is motivated by the fact that these devices constitute a large proportion of devices connected to edge networks. With access to much more sensitive information, smartphones and PCs are lucrative targets for attackers. Attackers exploit vulnerable IoT devices and use them to compromise high-end devices containing sensitive user data. Therefore, it is important to study the communications between single and multi-purpose IoT devices to detect any attacks.

\begin{figure}[th]
  \begin{center}
    \includegraphics[width=0.85\linewidth, bb=0 0 850 850]{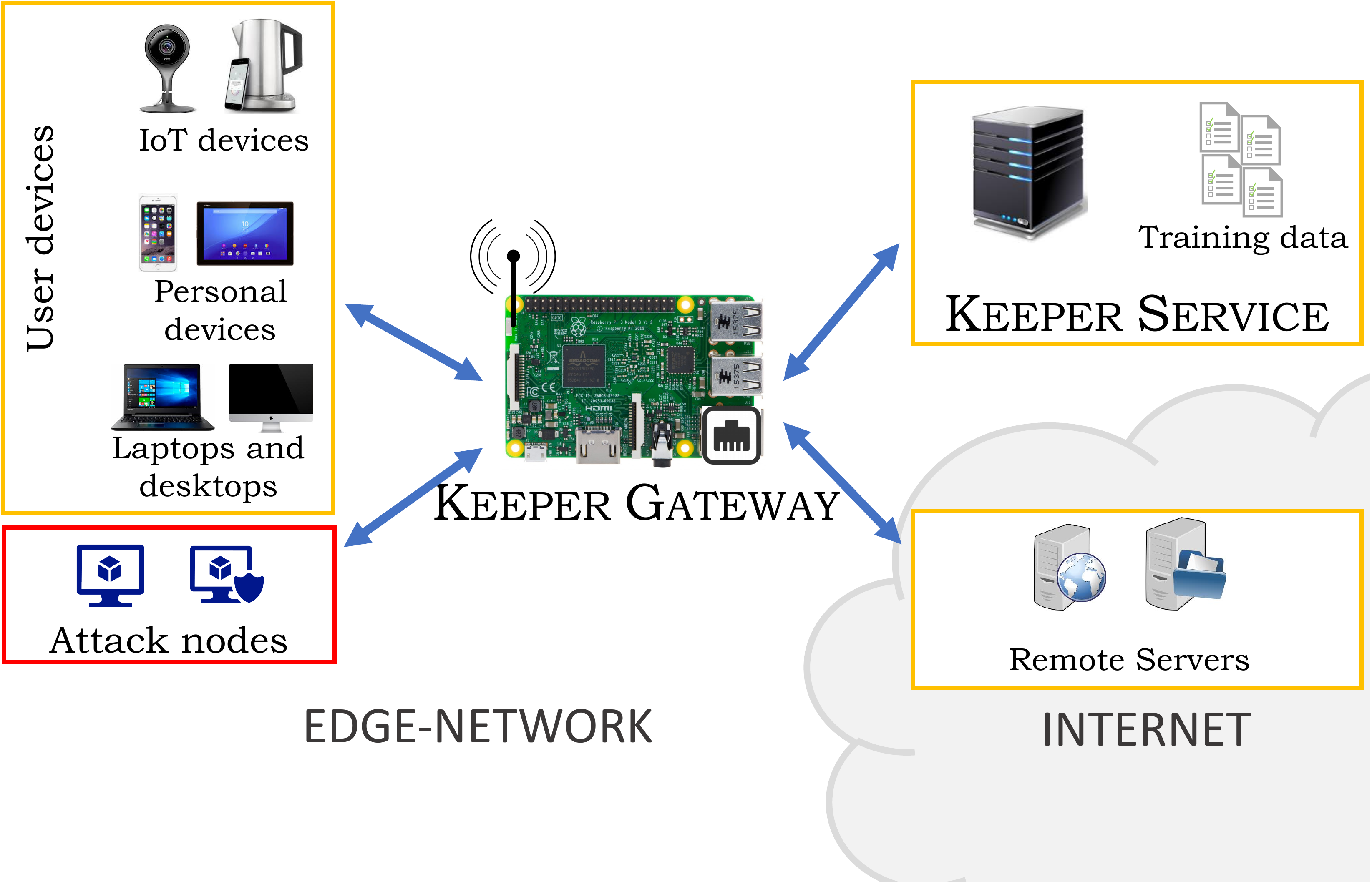}
  \end{center}
  \caption{Testbed used for data collection and system evaluation. \frontend is set up using Raspberry PI and operates as an access point, where user IoT devices and attacks nodes are connected. \backend is deployed using a consumer grade laptop.}
  \label{fig:testbed}
\end{figure}

\subsubsection*{Testbed Setup}
The network setup used for data collection is shown in Fig.~\ref{fig:testbed}. In this setup, all devices were connected to \frontend, which is deployed using a Raspberry-PI (\rpi), and setup as a wireless access point using \texttt{hostapd}. It also runs a \texttt{DHCP} server and manages \texttt{NAT} for both wired and wireless network. Current setup uses public \texttt{DNS} server but a local \texttt{DNS} can also be set up. 
All incoming and outgoing traffic from both wired and wireless interfaces is collected using \texttt{tcpdump}. Any traffic filtering was performed using layer-2 addresses. 
During data collection, \rpi was configured to drop all unfiltered outgoing traffic to prevent spread of \classM traffic on public Internet.

As mentioned before, all devices in the testbed support \wifi or wired connectivity. In case any device communicates to Internet via an IoT hub, using Weave, ZigBee or similar protocols, its D2I communications are monitored by capturing the network traffic generated by IoT hub.

\subsubsection*{Data collection}
The data collection process spans three phases of device activity:

\begin{itemize}
\item
\textbf{Setup}: This phase covers the network activity of a device when it is setup by the user for the first time. 

\item
\textbf{Background}: This phase covers the network activity of a device during its normal operation, including the phase when it connects or disconnects from the network. The background activity may vary with the kind of device, for example, \textit{single-purpose} devices may only generate heartbeat or status update messages, whereas, \textit{multi-purpose} devices may periodically fetch application updates, generate notifications etc.

\item
\textbf{Activity}: This phase covers the network activity of a device when it is actively communicating with other entities in the network. The network traffic generated by device corresponds to user interactions or messages communicated with other devices.
The network activity during this phase varies with the functionalities available in device. For example,~Dlink power plugs support only on/off functions, whereas a security camera allows user to switch on/off video feed, change video quality.
\end{itemize}

\begin{table*}[t]
\begin{tabular*}{\linewidth}{p{0.11\textwidth}p{0.08\textwidth}p{0.1\textwidth}p{0.40\textwidth}p{0.09\textwidth}p{0.1\textwidth}}
\toprule
\textbf{Classification} & \textbf{Activity} & \textbf{Tool}    & \textbf{Description}                                                         & \textbf{Evaluation samples} & \textbf{Training samples} \\
\midrule
\multirow{3}{*}{Scanning}          & Port Scan                    & ZenMap, NMap                & Scanning network for open ports on different hosts in the network                       & 1243647                                & 292092                           \\  
                                   & Port Sweep                   & ZenMap, NMap                & Scanning all TCP/UDP ports on one or more target hosts                                  & 953684                                 & 238421                               \\
                                   & Address sweep                & ARPing, ARP scan, Skipfish  & Scanning all hosts on the network and service running on them                           & 824363                                 & 229173                               \\
Botnet                             & Mirai                        & Telnet                      & Find and infect devices by deploying Mirai malware                 & 1389672                                & 437418                               \\ 
MitM                               & ARP Poisoning                & Arpspoof, EtterCap          & Using ARP poisoning attack to capture LAN traffic                                       & 1706479                                & 416619                               \\ 
Privilege escalation               & Fuzzing                      & PowerFuzzer, WFuzz,  Python & Searching vulnerabilities in devices connected to the network                           & 2356842                                & 559211                               \\ 
Data Theft                         & Data hijacking               & Telnet                      & Gain privileged access to other hosts and download collected data.                      & 821468                                 & 195371                               \\ 
Malware                            & Malware injection            & Metasploit                  & Upload malware to target hosts                                                          & 1161347                                & 304336                               \\ 
Denial of Service                  & SYN Flooding                 & Python scapy, Hyenae        & Flood the target host with many SYN requests to block it from performing any other task & 2801145       & 699543\\ 
								   & SSL renegotiation			  & tls-dos						&	Flood the target with SSL renegotiation packets to disable its packet stream &	3084492	& 671123\\
\bottomrule            
\end{tabular*}
\caption{Types of network attacks executed by compromised and malicious devices.}
\label{tbl:scenarios}
\end{table*}

To collect device \textit{setup} phase traffic, we collected all traffic from the device itself, as well as the management device used to setup the device. The device was reset and booted from factory default state prior to every time it was setup. For most devices, the firmware was upgraded before concluding the setup.
The \textit{background} traffic was collected by setting up a device and leaving it in connected state for a given time interval. The duration of these intervals ranged from 10 minutes to 72 hours.\footnote{(10, 20, 30) minutes, (1, 2, 6, 10, 12, 24, 36, 72) hours} To collect data for device \textit{activity}, user repeatedly performed an action on the device over a period of time, with irregular wait intervals between repetitions. The data collection was performed for different types of actions supported by the device. During data collection, the management device was either connected to same network, as the IoT device itself, or a remote network. After every iteration of data collection activity, network setup was reset to recover virgin network state for subsequent iteration. 

We assume every device to be inherently benign, therefore, the traffic it generates during standby and normal user interaction is considered its benign network behavior. 
Table~\ref{tbl:scenarios} lists different types of network attacks used for collecting traffic traces of malicious network activity. These attacks are commonly observed in IoT and edge networks~\cite{DBLP:journals/corr/MendezPY17, DBLP:journals/corr/abs-1708-05044, 7809147, NortonSmartHomeSecurity2017, KasperskyIoTScanner}. 

Table~\ref{tbl:scenarios} gives a high level classification and description for different types of attacks. It also lists the tools which are used to simulate these attacks. The number of training samples indicate the traffic flows used for model generation, whereas evaluation samples show the number of traffic flows used for evaluation. 
In order to emulate real world deployments, the volume of traffic handled by \sysname during evaluation is much higher compared to the volume of traffic used to train the system.
In addition to data collected from the testbed, we also use publicly available datasets for malicious IoT traffic~\cite{KitsuneDataset}.

\section{Evaluation}
\label{sec:evaluation}

\subsection{System Implementation}
\label{sec:implementation}
The evaluation testbed uses a Raspberry PI model~3 to deploy \frontend and a Core-i5 machine with 32GB memory to deploy \backend. \frontend runs Open vSwitch (\ovs)~\cite{188960} and Floodlight~\cite{FloodlightController} based SDN controller, with self-implemented custom modules that are used to perform traffic monitoring, traffic filtering, state management, \policy enforcement and cache management. The feature engineering and anomaly detection schemes were implemented with Python.
Both \frontend and \backend use REST-APIs for communicating with each other. 

\frontend was setup as a \wifi AP using \texttt{hostapd} module~\cite{Raspberry-as-AP}. All wired and wireless interfaces are bridged to \ovs, so that all network traffic is managed by \controller. In larger deployments, multiple OF switches are configured to use SDN controller running at the \frontend, for traffic management.
During evaluation, no data was sent to \backend and the initial classification model was trained using previously collected traffic data from the testbed.

This testbed setup serves as a reference implementation of \sysname. Our implementation was not optimized for performance gains. Therefore, the system and network performance results may vary with different hardware and software stacks used for system implementation. 

\subsection{Anomaly Detection}
\label{sec:eval_detection}
We studied the performance of anomaly detection technique in terms of sensitivity and false positive rates (FPR). Sensitivity, also known as recall, gives a measure of reliability of our technique in correctly identifying the malicious traffic flows, whereas, FPR gives an estimation of false alarms raised by the system, when benign activity is flagged as malicious. Ideally, the FPR rate should be zero, producing no false alarms. The trade-off between FPR and false negative rate (FNR) may vary with different scenarios. In general, low FPR may be preferred as it improves user experience by preventing false alarms. However, highly sensitive installations may require low FNR, so that no malicious traffic goes undetected and compromise the whole network. 
Using \sysname, false positives do not significantly impact user experience because \sysname enforces (and removes) network restrictions for any device, while maintaining minimal network access for the device, allowing it to continue its normal operations. This enables us to target lower FNR as well, for better security, without negatively affecting user experience.

We consider two types of classification problems:
\begin{itemize}
\item
\textit{Binary-class problem}: Differentiating between \classB and \classM network activity to detect anomalies.
\item
\textit{Multi-class problem}: Identify the sub-type of \classM activity exhibited by the device.
\end{itemize}

The motivation to identify the sub-type of malicious activity is that it provides us more information that can be used to enforce different levels of network restrictions for any device. For example, a device executing a network scanning attack may only be allowed to access its respective cloud service, whereas, network access for a device stealing user data should be completely blocked. This paper does not focus on identifying the sub-types of benign activity.

Our evaluation shows that \sysname was able to achieve an accuracy of $0.982$ with FPR$=0.01$ and FNR$=0.02$ for binary-class problem. It shows that our anomaly detection technique can differentiate between \classB and \classM network activity with high sensitivity. Based on these results, it can be concluded that \sysname is able to successfully identify block any malicious activity in the network. 

Table~\ref{tbl:multi_class} shows the performance achieved for identifying different types of \classM traffic. The results show that \sysname can identify volumetric attack (generating large volumes of traffic) with high sensitivity ($0.99$) and low FPR ($=0.02$). The network scanning attacks, in general, are detected with an accuracy of $0.993$ and f1-score $0.986$. This performance is better than the performance achieved for identifying different variants of network scanning attacks. 

In order to investigate this discrepancy, we study the feature value distributions in clusters representing these attacks. The feature value distributions represent the network behavior for different types of network activity. Therefore, if multiple network attacks have similar network footprint, the feature value distributions, observed in the clusters representing that traffic, will be overlapping. This overlap will result in misclassification. This phenomenon is prominent when we study different variants of network scanning attacks and it explains the relatively lower accuracies achieved for detecting variants of network scanning attacks. However, it should be noted that a network scanning attack, if it happens, is only misclassified as another network scanning attack. Since, the network restrictions for a device performing any type of network scanning attack are similar, the resulting security implications of these misclassification are negligible for given problem scenario.

Compared to volumetric attacks, it is difficult to detect detect MitM and data theft attacks because the network activity for these attacks is sporadic and difficult to distinguish from \classB traffic. However, \sysname achieves good performance in detecting these attacks, which otherwise go undetected by anomaly detection systems.

Our analysis revealed that device logs can also be useful for identifying the sub-type of \classM traffic. For example, analyzing network traffic generated during \textit{fuzzing} attack may register it as a \textit{network scanning} or DoS attack. However, studying device logs reveals it was a fuzzing attack against particular service running on target host.

\begin{table}[h]
\begin{tabular}{p{0.28\linewidth}llll}
\toprule
\textbf{Type} 	& \textbf{Accuracy} & \textbf{Recall} & \textbf{FPR} & \textbf{f1} \\
\midrule
Port Scan                         & 0.96             & 0.98                	& 0.07                	& 0.97             \\
Port sweep                        & 0.97             & 0.99                	& 0.06                 	& 0.98             \\
Address sweep                     & 0.97             & 0.99                	& 0.08                 	& 0.98             \\
Botnet                            & 0.99             & 0.99                	& 0.02                	& 0.99             \\
MitM                              & 0.77             & 0.92                	& 0.52                	& 0.85             \\
Fuzzing                           & 0.99             & 0.99                	& 0.01                	& 0.99             \\
Data theft                        & 0.74             & 0.88               	& 0.45                	& 0.77             \\
Malware injection                 & 0.79             & 0.94                	& 0.49                	& 0.88             \\
SYN flooding                      & 0.98             & 0.98                 & 0.03                	& 0.99              \\
SSL renegotiation                 & 0.96             & 0.99                 & 0.07                	& 0.97            \\ 
\bottomrule
\end{tabular}
\caption{Performance achieved by \sysname for identifying network attacks}
\label{tbl:multi_class}
\end{table}

\subsection{Network Performance}
\label{sec:eval_net_perf}
In order to maximize usability, it is vital for network security solutions to have minimal impact on user experience in terms of latency. Therefore, the design of \sysname is driven by the goal to minimize the latency experienced by end users. 

To study the impact on latency while browsing Internet, we studied page load times for top 1000 websites, ranked by Majestic~\cite{MajesticMillion}. The measurements were taken for three different scenarios including;

\begin{enumerate}
\item \sysname disabled.
\item \sysname enabled with $0\%$ cache hit rate.
\item \sysname enabled with $95\%$ cache hit rate.
\end{enumerate}

We compare the latency experienced when \sysname is disabled and no analysis is performed to the latency experienced when \sysname is enabled with anomaly detection and traffic filtering. $0\%$ cache hit rate means that the \policy cache is empty and all traffic flows are analyzed, whereas, $95\%$ cache hit rate means that $95\%$ of network traffic should have a matching policy available in the cache. Section~\ref{sec:sys_design} explains how caching reduces the number of requests made to perform anomaly detection, thereby, reducing the latency experienced by user as well as resource consumption. 

\begin{figure}[th]
\begin{minipage}{1\linewidth}
\centering
\subfloat[Page load times for top 1000 websites ranked by Majestic.]{\label{fig:page_load_times}\includegraphics[width=0.9\linewidth]{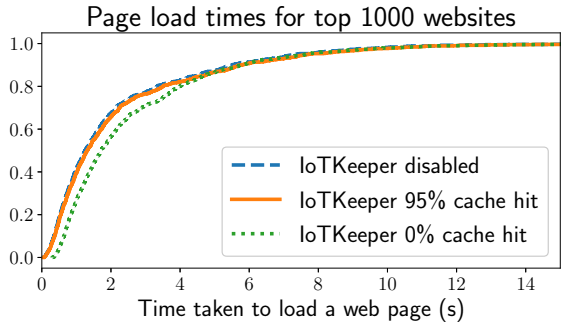}}
\end{minipage}

\begin{minipage}{1\linewidth}
\centering
\subfloat[Percentage of page load time consumed for traffic classification.]{\label{fig:prediction_overhead}\includegraphics[width=0.95\linewidth]{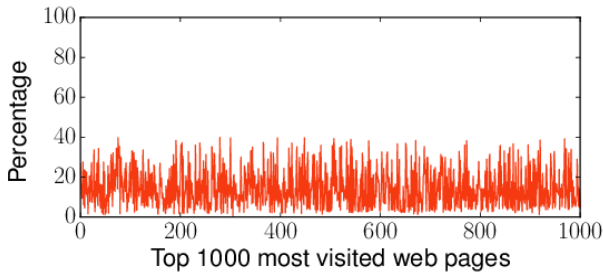}}
\end{minipage}

\caption{Impact of traffic classification over the latency experienced during web browsing.}
\label{fig:fis}
\end{figure}

Figure~\ref{fig:page_load_times} shows that when \sysname is enabled, average page load time is increased by upto $4.76\%$ and $15.89\%$ for $95\%$ and $0\%$ cache hit rate respectively. This increase in latency due to anomaly detection is not significantly high, even if there are no cached \policies available at \frontend. Investigating the increase in page load time for $0\%$ cache hit reveals that the relative increase in page load time is higher (up to $40\%$) for websites with very small page load times such as,~\url{google.com} and \url{microsoft.com}, but very low $\leq 7\%$ for websites with longer page load time such as,~\url{linkedin.com}, \url{instagram.com}. This is because the page load time for sites such as~\url{google.com} is very low ($\approx 0.5s$) and addition of constant time interval (required to perform analysis) will result in a large percentage increase for total page load time. On the other hand, this additional time will account for only a small percentage of time required to load websites with large page load times $\approx 2s$, such as,~\url{instagram.com},~\url{qq.com}.

The latency overhead does not depend on the volume of data loaded for webpage, instead it only depends on the time taken to identify traffic type and install flow table entry to handle the traffic. This overhead is only seen for the first time a web page is requested and any future requests for same web page will be handled by the \policy cache, with nearly no delay. Our experiments showed that with $100\%$ cache hit, latency is increased by $1.8\% \; (\pm1.49\%)$ only. 

Identifying the type of traffic flow is the most time consuming task performed by \frontend.
Figure~\ref{fig:prediction_overhead} shows the time taken to identify
a traffic flow type can account for $13.93\% \pm (9.55\%)$ percent of
the total time required to fetch a web page. In comparison, the time
taken for feature extraction, cache lookup and installation of flow
table rules is negligible. 


\sysname architecture suggests that \frontend will serve as a regular gateway used to setup edge networks. Therefore, we study the network performance achieved using \frontend, in detail. For this purpose, we measured the layer-4, layer-7 goodput, bufferbloat latencies as well as TCP and UDP latencies. 
Layer-4 goodput was calculated using \texttt{iperf3}\footnote{\url{https://iperf.fr/}} and layer-7 goodput was calculated for bulk file transfer using \texttt{curl}, to include protocol processing overhead as well.
Bufferbloat latencies was calculated using \texttt{netperf}\footnote{\url{https://github.com/HewlettPackard/netperf}} with \textit{RRUL test} (simulated by \textit{netperfrunner}~\footnote{\url{https://github.com/richb-hanover/CeroWrtScripts/blob/master/netperfrunner.sh}}) and speedtest (simulated using \textit{betterspeedtest}\footnote{\url{https://github.com/richb-hanover/CeroWrtScripts/blob/master/betterspeedtest.sh}}). These tests use multiple simultaneous connections to simulate heavy network load to study latency and throughput in uplink and downlink.
Lastly, TCP and UDP latencies were calculated using \texttt{qperf}\footnote{\url{https://linux.die.net/man/1/qperf}}.

The experiments were conducted to study the performance for D2D (LAN$\leftrightarrow$LAN) and D2I (LAN$\leftrightarrow$WAN) communications. For each type, we compared the performance achieved in \textit{insecure} setting with \sysname disabled and \textit{secure} setting with \sysname enabled for anomaly detection and traffic filtering.

\begin{table*}[]
\begin{center}
\begin{tabular}{p{0.17\linewidth}lllll}
                        &                    & \multicolumn{2}{c}{D2D}                                         & \multicolumn{2}{c}{D2I}                                         \\
\textbf{Metric}         & \textbf{Direction} & \textbf{Insecure} & 	\textbf{Secure} & \textbf{Insecure} & \textbf{Secure} \\
Layer 4 goodput              & Up                 & $89.97\;(\pm 0.77)$                 & $89.69\;(\pm 0.03)$                   & $90.11\;(\pm 0.80)$                   & $88.91\;(\pm 0.10)$                   \\
                        &                    & $90.46\;(\pm 0.34)$                 & $89.70\;(\pm 0.02)$                  & $91.01\;(\pm 1.53)$                   & $89.70\;(\pm 0.15)$                  \\
Layer 7 goodput              & Up                 & $87.67\;(\pm 1.32)$                 & $84.152\;(\pm 0.12)$                 & $89.94\;(\pm0.60)$                 & $86.23\;(\pm 0.34)$                 \\
                        & Down               & $88.60\;(\pm1.52)$                  & $88.17\;(\pm 2.42)$                  & $89.12\;(\pm0.89)$                 & $87.78\;(\pm 1.22) $                  \\
Bufferbloat latency (ms) (Speedtest) & Up                 & $2.11\;(\pm 0.40)$                   & $3.02\;(\pm 0.36)$                  & $3.77\;(\pm 0.24)$                  & $3.01\; (\pm 0.36)$                  \\
                        & Down               & $90.71\;(\pm 2.01)$                  & $92.02\;(\pm2.31) $                 & $81.41\;(\pm 2.67)$                 & $82.83\;(\pm2.10) $                  \\
Bufferbloat latency (ms) (RRUL test)     & Up                 & $2.11\;(\pm 0.13)$                   & $2.82\;(\pm 0.44) $                  & $2.92\;(\pm 0.89)$                  & $3.22\;(\pm 0.77)$                  \\
                        & Down               & $45.81\;(\pm 1.73)$                 & $50.13\;(\pm 1.44)$                  & $54.11\;(\pm 1.87)$                & $55.93\;(\pm 2.44)$                  \\
Latency (ms)                & TCP                & $ 0.37\;(\pm 0.004)$                & $0.42 (\pm 0.003)$                & $0.38\;(\pm 0.003)$                 & $0.38\;(\pm 0.004)$                \\
                        & UDP                & $0.38\;(\pm 0.003)$                 & $0.40\;(\pm 0.003)$                & $ 0.39\;(\pm 0.004)$                & $0.39\;(\pm 0.003)$               
\end{tabular}
\end{center}
\caption{Network performance achieved by \sysname, in terms of throughput and latency, with \rpi based deployment}
\label{tbl:net_perf}
\end{table*}

These results give a qualitative and quantitative understanding of the overheads on network performance due to traffic classification. It is evident that \sysname does not introduce significant deterioration in network performance in comparison with baseline performance using same hardware.

\subsection{System Performance}
\label{sec:eval_sys_perf}
We investigated the deployment feasibility of \sysname, using a \rpi, for analyzing network traffic at line speeds. 
The experimentation was conducted using traffic traces collected for duration of 30 and 60 minutes respectively, from a fully saturated link on \rpi. 
We calculated the time required to analyze all data points in these samples, using different number of clusters $i$.

\begin{figure}
\centering
\includegraphics[width=\linewidth]{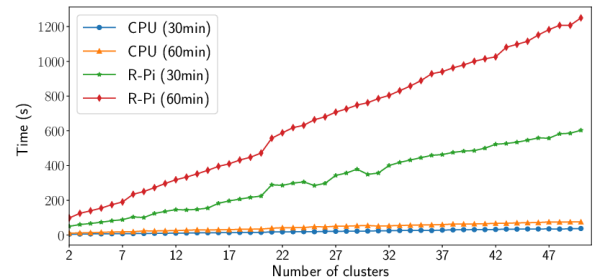}
\caption{Time required for analyzing traffic samples, using different number of clusters}
\label{fig:cluster_times}
\end{figure}

Figure~\ref{fig:cluster_times} shows that the time required to cluster and analyze the data points increases linearly as the number of clusters increase. 
It can be observed that for $i=17$, all data points in 30 minutes sample can be analyzed in less than 4 minutes using a \rpi. For comparison, same analysis takes less than 30 seconds on a consumer grade Core i5 laptop with 32Gb memory. Similarly, we can analyze the 60 minute sample, with $i=50$, in approximately $21$ minutes, using \rpi. 
These results show that \sysname is able to operate at line speeds using low-cost single board computers. 

\begin{figure}[h]
\begin{minipage}{1\linewidth}
\centering
\subfloat[Relationship between the number of cached \policies and the number of classification requests made for handling new traffic flows.]{\label{fig:cache_vs_policies}\includegraphics[width=0.8\linewidth]{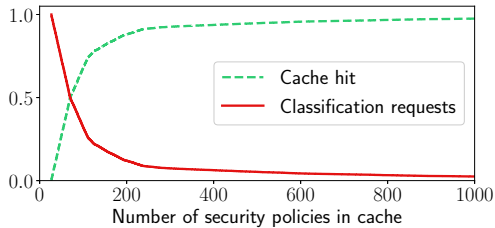}}
\end{minipage}

\begin{minipage}{1\linewidth}
\centering
\subfloat[Behavior of cache lookup time and (deep memory) cache size relative to the number of security policies stored locally.]{\label{fig:cache_lookup}\includegraphics[width=0.8\linewidth]{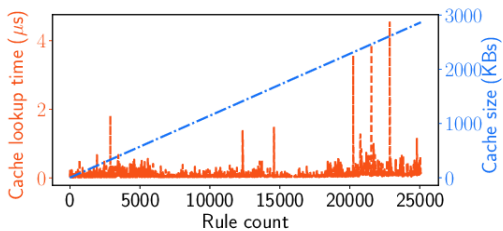}}
\end{minipage}
\caption{Effects of cached \policies over system performance}
\label{fig:cache_performance}
\end{figure}

In order to study the impact of caching over the system performance, we studied how the number of \policies in cache affects the number of classification operations performed. Figure~\ref{fig:cache_vs_policies} shows that $90\%$ traffic flows in the network can be handled using roughly $200$ \policies in the cache. These results validate the hypothesis that most of network traffic from edge networks is destined to only few cloud services. Therefore, a relatively small number of cached \policies can result in high cache hit rate, thereby, lowering the latency, as well as resource footprint. 

We analyze our hash-table based implementation of cache to study the performance in terms of lookup time and cache size, relative to the number of cached \policies. As expected, the deep-memory size of cache increases linearly with number of \policies while lookup time remains constant. A few spikes in lookup time can be attributed to underlying hardware and operating system. Section~\ref{sec:design_frontend} discusses how we can limit the linearly increasing cache size using expiry time and setting an upper bound on the maximum  disk space available for cache.

\section{Related Work}
\label{sec:related_work}
A number of techniques have been proposed to identify anomalies in network traffic~\cite{4738466, 7346821, 6510197, Shon2005, nids-fuzzy-2011, Bohara:2016:IDE:2898375.2898400}. Researchers have studied various feature analysis and machine learning techniques to identify botnets~\cite{Gu:2008:BCA:1496711.1496721, Strayer2008, Lu:2009:ADB:1533057.1533062}, Denial of Service~\cite{6510197} and other attacks in the network~\cite{766909,IoTurva:MobiCom17}. 

The anomaly detection techniques can be categorized into two types: \textit{offline} and \textit{online}.
Offline techniques are developed using labeled dataset. Offline algorithms have access to whole dataset and multiple iterations of training and evaluation are performed to produce the final classification model. This exercise consumes lots of time and resources. These algorithms are often used by signature-based network security solutions to identify network attacks. 

Online techniques do not have access to complete dataset during model training and they mostly use unlabeled data to perform network traffic classification. They also need to be efficient enough to ensure high detection accuracy at high packet arrival rates, using limited resources. 

Majority of existing traffic anomaly detection techniques are developed for offline analysis~\cite{4738466, Gu:2008:BCA:1496711.1496721} and require labeled data. Given the number and variety of IoT devices, data collection is challenging. Crowd-sourced data collection has been proposed to address the problem~\cite{DBLP:journals/corr/abs-1804-07474} but it has its own limitations~\cite{7384520}. 

\sysname is an online anomaly detection technique. Therefore, we do not compare its functioning and performance with signature-based network security solutions as these solutions are highly specialized, operate on custom hardware, incur high deployment and operational costs. The performance of signature-based solutions is also limited by the availability of attack signatures.

Among online traffic classification techniques, Securebox was, to the authors' knowledge, the first one to propose a two tier model, where a lightweight network gateway uses a cloud service to analyze traffic from edge networks~\cite{Hafeez:2016:STS:3010079.3012014}. The cloud service supports traffic analysis using software middleboxes, and various machine learning based analysis technique. \sysname addresses the privacy and latency problems in Securebox model by performing traffic analysis locally on the network gateway. 

IoT Sentinel~\cite{7980167} identifies the IoT devices by analyzing their network traffic and sets up network access control based on the profile of given IoT device. While IoT Sentinel detects device types with high accuracy, it requires network traces captured during device setup, for this purpose. If a device was already setup, before it was connected to network, IoT Sentinel unable to identify the device and fail to setup network restrictions. Also, the access control policies are coupled with IoT devices and IoT Sentinel does not provide a mechanism to update these policies with the evolution of devices' network behavior. In comparison, \sysname constantly monitors network activity to detect and block malicious traffic, at any time, irrespective of what device is generating the traffic.

Recently proposed Kitsune~\cite{DBLP:journals/corr/abs-1802-09089} uses an ensemble of autoencoders for high accuracy online anomaly detection. Kitsune uses incremental damped statistics to extract features and track devices' network behavior. To reduce memory footprint, Kitsune maintains information about device behavior for a fixed time. Hence, if a device exhibits anomalous behavior long enough, the model will consider it as \textit{normal} behavior. 
To address this issue, \sysname maintains device behavior information through the timeline of its connectivity. This information is used to compare devices' latest behavior to its previous behavior at any point in time. It enables us to detect changes in devices' network behavior due to firmware updates and configuration changes.

DIoT~\cite{DBLP:journals/corr/abs-1804-07474} uses the periodicity in IoT device traffic to identify device type and uses device-type-specific anomaly detection model to detect network attacks. Although this technique achieves high accuracy, the anomaly detection model depends on device type identification. 
Given the huge variety of devices, it is difficult to develop and maintain device-type-specific anomaly model. Meanwhile, any wrong device type identification results will essentially render the device useless, thereby, negatively affecting user experience. DIoT also does not update the anomaly detection model based on changes in device configuration and software updates. Compared to DIoT, \sysname does not require device-type information to detect anomalies and the anomaly detection scheme can also accommodate any changes in network behavior due to changes in device configuration.

Anomaly detection techniques such as, DIoT, BotMiner~\cite{Gu:2008:BCA:1496711.1496721} mainly detect volumetric attacks (producing large volume of network traffic) such as, Mirai botnet. It is difficult for these techniques to detect attacks such as MitM, ARP Spoofing, which have sporadic network activity similar to normal device activity. \sysname, on the other hand, is able to detect both these kinds of attacks with high accuracy.

Online detection techniques~\cite{7980167, DBLP:journals/corr/abs-1802-09089, DBLP:journals/corr/abs-1804-07474} use the intrinsic device behavior as its normal behavior. Hence, they are unable to detect any anomalies in devices' network behavior if it is inherently compromised. Meanwhile, \sysname can capture any discrepancies in a devices' \textit{normal} network behavior, during clustering, and labels such behavior as malicious, depending on the feature value distributions observed in given cluster. Therefore, we are able to detect malicious behavior of inherently compromised devices.  

Recently proposed anomaly detection techniques also use \textit{recurrent neural networks}~\cite{Du:2017:DAD:3133956.3134015, DBLP:journals/corr/abs-1805-03735, Malhotra2015LongST, DBLP:journals/corr/BontempsCML17} or \textit{gated recurrent units}~\cite{DBLP:journals/corr/abs-1802-00324, 7266837} for anomaly detection. Some techniques model network traffic as \textit{symbols} in a language and use a frequency based model to identify anomalous sequence of symbols, indicating network anomalies~\cite{DBLP:journals/corr/abs-1805-03735, DBLP:journals/corr/abs-1804-07474}. These techniques are mainly employed for offline analysis and have high resource footprint.

Any technique which performs remote traffic analysis raise security and privacy concerns for users whose traffic data is analyzed in remote environment. The data storage, processing and analysis in these environments is beyond users' control and the traffic data being analyzed contains sensitive user data and personally identifiable information. Using \sysname, we alleviate these concerns by performing traffic analysis within user network, where user has complete ownership and control over the data being analyzed by the network.

Software middleboxes are proposed for on-demand traffic analysis using cloud infrastructure~\cite{Sherry:2012:MMS:2377677.2377680}. Middlebox virtualization reduces deployment costs and improves scalability. However, the increase in latency experienced by re-routing traffic through these middleboxes, cost of analyzing huge volumes of network traffic, and privacy implications of analyzing business critical data under third party control, are some of the challenges faced by these proposals.

Various commercial products such as, Cujo\footnote{\url{https://www.getcujo.com/smart-firewall-cujo/}}, Dojo\footnote{\url{https://dojo.bullguard.com/}}, Core\footnote{\url{https://us.norton.com/core}}, Sense\footnote{\url{https://sense.f-secure.com/}} have been launched to protect IoT and smart homes. These products claim to perform real-time behavioral traffic analysis and deep packet inspection to detect network attacks. At the time of writing, not all of these features are available on latest generation of these devices. Due to limited resources on gateway, most products perform traffic analysis in their respective cloud services, raising aforementioned privacy challenges. Ensuring low latency and high network throughput performance is also a big challenge for these products, which essentially use a proxy to intercept and analyze user traffic flowing through the gateway, impacting the latency and throughput. 

Some of these devices claim to perform deep packet inspection on the router itself, resulting in severely degraded network performance. The growing use of encrypted protocols also limits the usefulness of deep packet inspection. Since \sysname does not perform payload analysis, its performance is not limited by the use of encrypted protocols. 

\section{Discussion}
\label{sec:discussion}
\sysname is a network-based security solution designed to detect anomalies and react to those by isolating the devices exhibiting malicious behavior. Our evaluation demonstrates that the proposed solution efficiently secures edge networks against any attacks. We now discuss possible shortcomings and limitations of \sysname. 

\subsection*{Feature engineering and model generalization}
Feature engineering is an important step in developing a generalized detection model. The resource limitations of single-board computers required us to identify the least number of features capturing maximum variance in the network traffic data, to detect anomalies. The final feature set had to be compact and generalizable such that it results in consistent anomaly detection performance across multiple datasets. To address these requirements, we analyzed each feature individually to study its significance for traffic classification.
Our analysis revealed that a concise feature set, extracted from network data, can successfully identify anomalies in network traffic. Meanwhile, device logs, if available, can also be helpful in improving the performance of anomaly detection scheme.

It should be noted that \sysname does not perform deep packet inspection or use any features extracted from unencrypted payload analysis. It can identify malicious network activity of any connected device but cannot detect any malicious data included in packet payload.

\subsection*{Detecting non-volumetric attacks}
The network activity of volumetric attacks such as denial-of-service attacks, is substantially different from regular device activity because of high traffic volumes and protocols used. We can achieve high accuracy in detecting volumetric attacks, using features extracted from traffic metadata only. However, it is not possible to achieve similar performance, using same feature set, if the network footprint of an attack is small and infrequent such as, MitM attacks. 

Although \sysname is able to detect these attacks with infrequent network activity, the performance of anomaly detection can be improved by using human expertise to analyze the underlying model. In this regard, \backend can collect various statistics about classification models trained by \frontend and human experts can analyze this data to identify any discrepancies in the anomaly detection model. Any updates, if needed, are sent from \backend to all gateways deployed in edge networks.

\textbf{Free loaders}:
By default, \sysname restricts the network access for malicious devices but it does not fully block their access to the Internet. Although this strategy prevents any attacks in the network, it does not block free loaders from consuming network bandwidth. However, \frontend allows users to monitor and limit the bandwidth consumed by connected devices, to prevent these free loaders from exhausting limited bandwidth. 
 
\subsection*{Evolution in device behavior}
The ability of \sysname to identify device firmware upgrades and configuration changes allows us to limit number of false alarms raised by the system, as well as track the progress of software updates for device deployed in the wild. The knowledge of firmware versions (operated by IoT devices) allows us to readily update \policies, to prevent any attempts to exploit known issues and vulnerabilities in the given firmware version. 
Given that \sysname can identify these updates, it does not detect minor upgrades such as, software patches, which do not have significant impact on device' network behavior.

\textbf{Physical tampering with devices}:
\sysname monitors network traffic to detect any malicious activity. Hence, it is not able to detect if a device has any backdoors or is physical tampered with, by an adversary. We assume that any backdoors or physical tampering are motivated by malicious intent to influence devices' behavior to the favor of adversary. Since majority of IoT devices are connected to the network, any malicious behavior will be detected and blocked by \sysname.

\textbf{Cellular and bluetooth communications}:
\sysname only monitors the communications passing through \frontend. Any communications using other channels such as, cellular data, satellite link, can not be secured by the proposed system. 
The current implementation does not monitor D2D communications occurring via low-power communication protocols. Our study revealed that IoT devices do not generally use low-power protocols for D2D communications and such communications are performed via IoT hub, which can be monitored.

\subsection*{Attacks against \sysname}
Our system design limits the attack surface of \frontend by requiring physical proximity or access to cloud service to perform any configuration changes. In case if an adversary gains access to user credentials for the cloud service configuration portal, it can reset or disable security features on \frontend and render it useless. To prevent realization of such attacks, \sysname architecture supports the use of 2-factor authentication, notifications about configuration changes and ability to roll back changes to any point in time using state backups. 

\textbf{MAC address spoofing}:
\sysname sets up network access restrictions based on layer-2 MAC addresses. An adversary can circumvent these restrictions by spoofing device MAC address. 
In such scenarios, as long as the adversary does not exhibit malicious activity, it will have regular network access but this behavior has no incentive for the adversary. On the contrary, if adversary engages in malicious activity with spoofed MAC address, \sysname will identify and block that activity. 

\textbf{Denial of Service}:
There is a possibility that adversary can exploit MAC address spoofing to perform DoS attack against \frontend. In that case, caching will limit the number of times similar traffic flows, coming from different MAC addresses, are analyzed by the gateway. Moreover, our evaluation also shows that \sysname is able to perform anomaly detection at line speeds without becoming a bottleneck. An adversary can flood the upstream link in \frontend but this will block all traffic flows in the network, including attacker's own traffic, giving no incentive to the attacker. 

There is also a possibility of DDoS attacks against \backend. Due to the system architecture, these attacks do not affect \frontend functionality because it does not depend on \backend. Meanwhile, the attacks against \backend can be handled using several known techniques to prevent DDoS attacks. To prevent \backend from becoming single point of failure when issuing updates, peer-to-peer protocols with checksums and public-key encryption, can be used for transmitting updates to \frontend deployed in edge networks.

\textbf{Adversarial machine learning}:
An advanced adversary can use adversarial machine learning techniques~\cite{DBLP:journals/corr/MillerHQK17} to understand the anomaly detection model and generate specially crafted packet flows to circumvent detection mechanism. However, this approach is infeasible because even small changes in packet headers substantially change the network behavior. Meanwhile, payload modifications do not help in circumvention because \sysname does not perform any payload analysis. 
Therefore, it is difficult make small enough changes in packets' header space, which preserve the malicious intent and do not change characteristics of network flow. 

\subsection*{Scalability}
Current architecture uses \backend to only provide support services for \frontend but system architecture allows us to deploy additional analytics services in the \backend as well. For example, \frontend can re-route traffic from specific devices through middleboxes and perform sophisticated analysis. 
In order to speed up model training process, \frontend can use the resources available in \backend to offload model training. It is also possible to use the computational power of devices connected to the network, such as PC, smartphones, for training anomaly detection model and performing computationally intensive traffic analysis. 

\section{Conclusion}
\label{sec:conclusion}
This paper presents \sysname, a platform for securing edge networks by detecting malicious network traffic and isolating the devices generating that traffic. 
\sysname platform adopts lightweight design and can be deployed using low-cost programmable devices so that traffic classification and \policy enforcement can be performed at the network gateway, in real time.
It relies on the feature set extracted from network traffic data, to successfully identify various types of network attacks. 
The ability to dynamically generate and enforce \policies enables automation of network configuration and readily blocks any malicious actor in the network, using adhoc overlay networks.
\sysname evaluation, using a real world testbed, demonstrates that \sysname can successfully perform traffic analysis on network gateways, with minimal impact on user experience. Moreover, it does not require sophisticated hardware or modifications on existing IoT and other devices for its operations.


\bibliographystyle{elsart-num} 

\bibliography{references}

\end{document}